\documentclass[a4paper]{article}

\usepackage[utf8x]{inputenc}

\usepackage{amsmath}

\usepackage{amssymb}

\usepackage{graphicx}
\usepackage{subfigure}

\usepackage{microtype}

\usepackage{tabu}

\usepackage[english]{babel}

\begin{document}

\title{A Survey of Entorhinal Grid Cell Properties}

\author{Jochen Kerdels \and Gabriele Peters}

\maketitle              

\begin{center}
University of Hagen - Chair of Human-Computer Interaction\\
Universit\"{a}tsstrasse 1, 58097 Hagen - Germany
\end{center}

\begin{abstract}
\noindent About a decade ago {\em grid cells} were discovered in the medial 
entorhinal cortex of rat. Their peculiar firing patterns, which correlate with 
periodic locations in the environment, led to early hypothesis that grid cells 
may provide some form of metric for space. Subsequent research has since 
uncovered a wealth of new insights into the characteristics of grid cells and 
their neural neighborhood, the para\-hip\-po\-campal-hippocampal region, calling
for a revision and refinement of earlier grid cell models. This survey paper 
aims to provide a comprehensive summary of grid cell research published in the 
past decade. It focuses on the functional characteristics of grid cells such as
the influence of external cues or the alignment to environmental geometry, but 
also provides a basic overview of the underlying neural substrate. 
\end{abstract}

\section{Introduction}

The para\-hip\-po\-campal-hippocampal region (PHR-HF) of the mammalian brain 
hosts a variety of neurons whose activity correlates with a number of 
allocentric variables. For example, the activity of so-called {\em place} cells 
correlates with specific, mostly individual locations in the 
environment~\cite{OKeefe1971,OKeefe1976}, the activity of {\em head direction} 
cells correlates with the absolute head direction of an 
animal~\cite{Taube1990,Taube1995}, the activity of {\em grid} cells correlates 
with a regular lattice of allocentric locations~\cite{Fyhn2004,Hafting2005}, the
activity of {\em border} cells correlates with the proximity of an animal to 
specific borders in its environment~\cite{Solstad2008,Savelli2008}, and, 
finally, the activity of {\em speed} cells correlates with the current speed of 
an animal~\cite{Kropff2015}. In addition, there are a number of cells whose 
activity reflect a combination of those variables. Together, these cells are 
commonly assumed to be part of a system that supports tasks of spatial 
representation, orientation, and navigation. Within this system grid cells stand
out from the other neurons by their remarkable grid-like firing patterns that
resemble a set of carefully oriented rulers spanning the entire environment. 
Hence, they are often seen as providing some kind of {\em metric for space} to 
the animal~\cite{Moser2008a,Moser2008b}.

This survey aims to provide an overview of functional grid cell properties as 
well as information on the underlying neuronal structures present in a major 
area hosting grid cells, the medial entorhinal cortex (MEC). The survey is based
on previous work by one of the authors~\cite{Kerdels2016}, but is significantly 
extended and restructured to be comprehensive and self-contained. The survey is 
structured as follows. Section~\ref{sec:gmeasures} introduces key measures that 
are widely used to identify and characterize grid cells on a functional level.
Section~\ref{sec:development} and section~\ref{sec:toporg} report findings 
regarding the development of grid cells and their topographical organization, 
respectively. Section~\ref{sec:funcsec} constitutes the main part of this survey
and covers important functional grid cell properties that were discovered within
the last decade. Subsequently, section~\ref{sec:neurostruct} turns to the 
underlying neural substrate reporting on findings regarding the general 
structure of the MEC as well as specific findings regarding possible grid cell 
microcircuits. Finally, section~\ref{sec:summary} concludes this survey. 

\section{Grid Cell Measures}
\label{sec:gmeasures}

Grid cells stand out from other neurons in the PHR-HF by their triangular, 
grid-like firing patterns. To characterize the spatial properties of this grid 
structure Hafting et al.\@~\cite{Hafting2005} established four measures that are 
used throughout the grid cell literature: {\em spacing}, {\em orientation}, {\em
field size}, and {\em phase} of a grid cell. In addition, Sargolini et 
al.\@~\cite{Sargolini2006} introduced a {\em gridness} score which quantifies the
degree of spatial periodicity of a cell's firing pattern.

The basis of all five measures is the {\em firing rate map} of the grid cell in
question. The firing rate map of a cell is typically constructed by discretizing
the environment into bins of equal size (e.g. \(3\mathrm{cm} \times 
3\mathrm{cm}\)) and determining the spatially smoothed, average firing rate for 
each bin. For instance, Sargolini et al.\@~\cite{Sargolini2006} estimate the 
average firing rate \(\lambda\left(x\right)\) of the bin centered on position~\(x\) 
as:
\[
\lambda\!\left(x\right) = \sum_{i = 1}^{n}g\!\left(\frac{s_i - x}{h}\right) \bigg/
  \int_{0}^{T}g\!\left(\frac{y\left(t\right)-x}{h}\right)dt
\]
with a Gaussian kernel \(g\), a smoothing factor \(h = 3\), the number of spikes 
\(n\), the location \(s_i\) of the \(i\)-th spike, the location \(y\left(t\right)\) of
the rat at time \(t\), and the recording period \(\left[0,T\right]\). To avoid 
extrapolation errors bins further than the bin width apart from the tracked path
of the animal are considered as unvisited. In a more recent publication, 
Stensola et al.\@~\cite{Stensola2012} use a \(5 \times 5\) boxcar average instead
of a Gaussian kernel for smoothing. The use of this boxcar average results in
firing fields that appear more accentuated and crisp compared to the Gaussian 
smoothing.

Four of the five grid measures, i.e., spacing, orientation, field size, and 
gridness require the calculation of a spatial autocorrelogram of the grid cell's
firing rate map. Sargolini et al.\@~\cite{Sargolini2006} construct this 
autocorrelogram using the Pearson product-moment correlation coefficient 
\(r\!\left(\tau_x,\tau_y\right)\) to calculate the autocorrelation between rate 
map bins \(\lambda\) separated by \(\left(\tau_x,\tau_y\right)\):
\small
\[
\frac{
  n\sum \lambda\!\left(x,y\right)\lambda\!\left(x-\tau_x,y-\tau_y\right) - 
  \sum \lambda\!\left(x,y\right) \sum \lambda\!\left(x-\tau_x,y-\tau_y\right)
}
{
  \sqrt{n \sum \lambda\!\left(x,y\right)^2 - 
    \left(\sum \lambda\!\left(x,y\right)\right)^2}
  \sqrt{n \sum \lambda\!\left(x-\tau_x,y-\tau_y\right)^2 - 
    \left(\sum \lambda\!\left(x-\tau_x,y-\tau_y\right)\right)^2}
}
\] 
\normalsize
where the summation is over all \(n\) bins for which both \(\lambda\!\left(x,y
\right)\) and \(\lambda\!\left(x-\tau_x,\right.\) \(\left.y-\tau_y\right)\) have 
valid entries in the firing rate map. Autocorrelations for shifts where \(n < 
20\) are not included in the resulting autocorrelogram.

\begin{figure}[t]
    \begin{center}
        \subfigure[]{\includegraphics[width=0.2\textwidth]{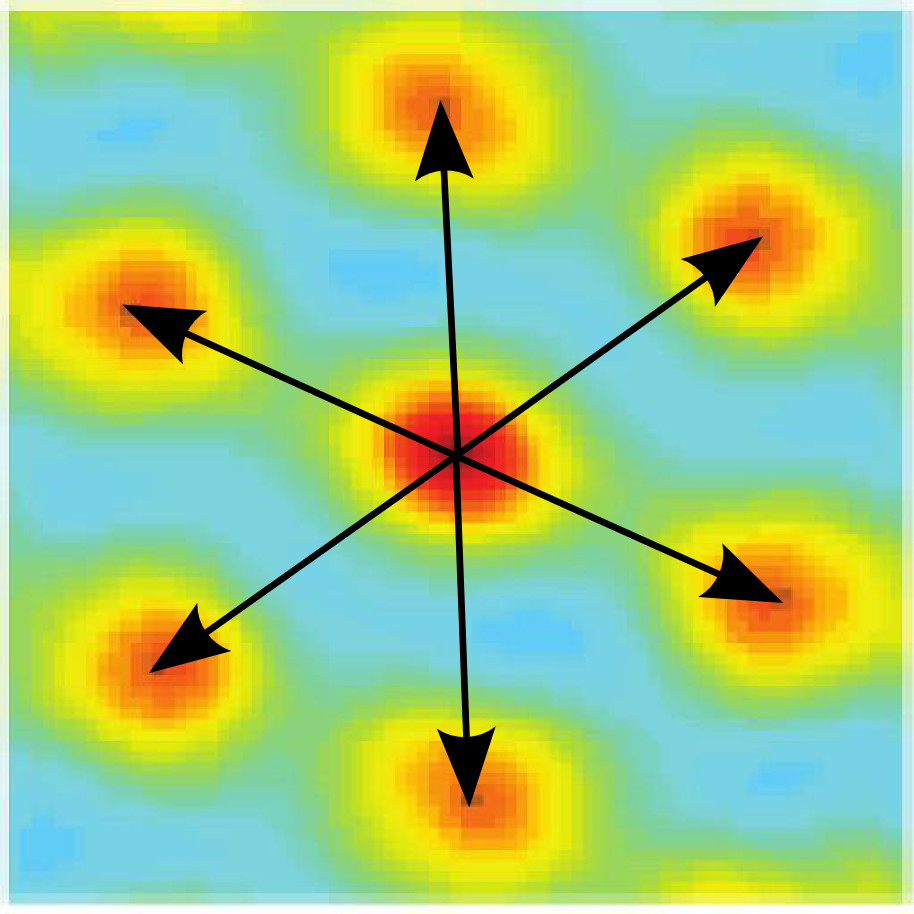}}\hspace{1.5cm}
        \subfigure[]{\includegraphics[width=0.2\textwidth]{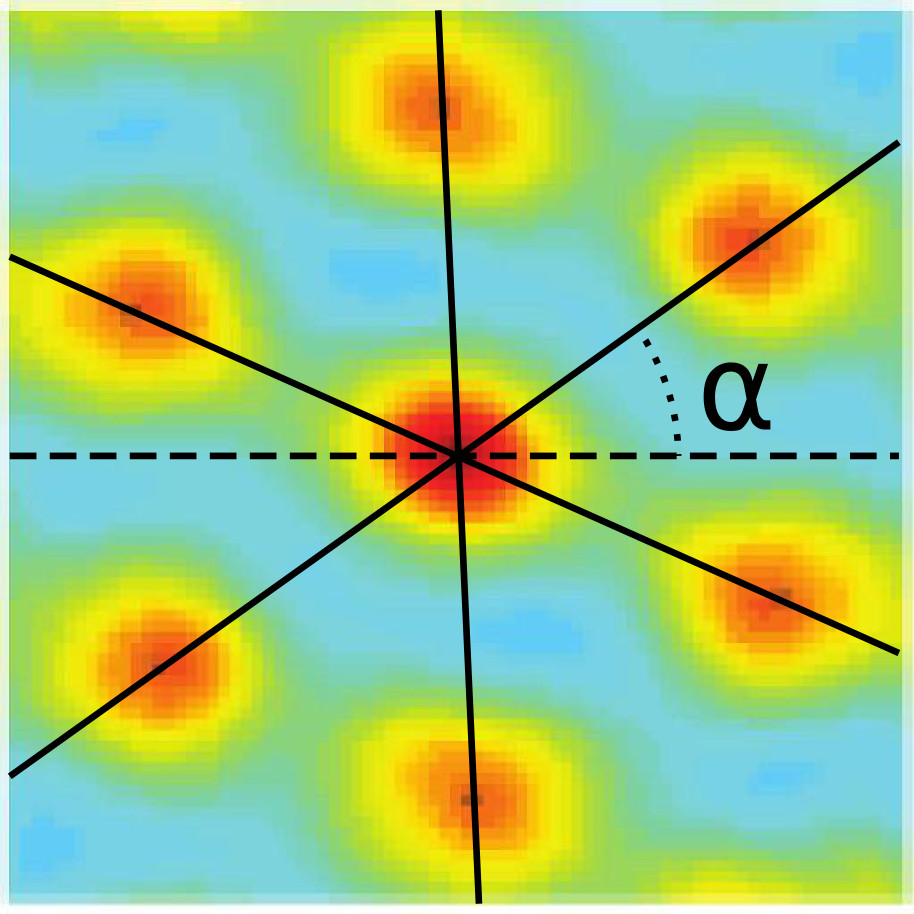}}
        \caption[grid spacing and orientation]{Grid cell spacing and 
          orientation.({\bf{a}}) The {\em spacing} of a grid cell is defined as 
          the median distance between the center peak and the six surrounding 
          peaks in the autocorrelogram. ({\bf{b}}) The {\em orientation} of a 
          grid cell is defined as the angle \(\alpha\) between a fixed reference 
          line (dashed) going through the central peak and the closest of the 
          three main diagonals of the grid cell hexagon in counterclockwise 
          direction.
          Figures based on autocorrelogram from Sargolini et al.\@~\cite{Sargolini2006}.}
        \label{fig:grid_spacor}
    \end{center}
\end{figure}

\begin{figure}[t]
  \centering
  \includegraphics[width=1.00\textwidth]{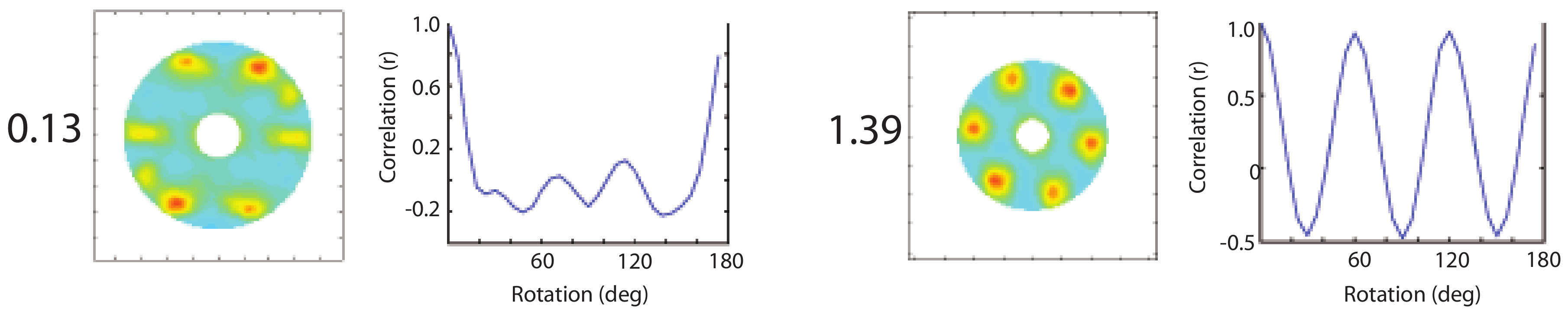}
  \caption[gridness]{Gridness scores of two grid cells. The shown 
    autocorrelograms include only the area containing the six peaks surrounding 
    the center that was used for the calculation of the gridness scores. The 
    graph to the right of each autocorrelogram shows the correlation of the
    particular autocorrelogram with a rotated version of itself in \(6^\circ\) 
    steps. Gridness scores are given to the left of each autocorrelogram.
    Figure adapted from Sargolini et al.\@~\cite{Sargolini2006}.}
  \label{fig:gridness}
\end{figure}

Based on the autocorrelogram of the grid cell's firing rate map {\em spacing},
{\em orientation}, {\em field size}, and {\em gridness} are defined as follows.
The {\em spacing} of a grid cell (fig.~\ref{fig:grid_spacor}a) is defined as the
median distance between the central peak of the autocorrelogram and its six 
surrounding peaks. The {\em orientation} of a grid cell 
(fig.~\ref{fig:grid_spacor}b) is defined as the angle between a fixed reference 
line (0~degrees) going through the central peak of the autocorrelogram and the 
closest of the three main diagonals of the surrounding hexagon in 
counterclockwise direction. The {\em field size} of a grid cell refers to the 
size of the individual firing fields. It is estimated as the area occupied by 
the central peak in the autocorrelogram with respect to a fixed threshold, e.g.,
\(r = 0.2\) as used by Hafting et al.\@~\cite{Hafting2005}. To calculate the {\em 
gridness score} of a grid cell only the six peaks surrounding the central peak 
in the autocorrelogram are taken into account. All other regions of the 
autocorrelogram including the central peak are masked out. Then, the correlation
values between the masked autocorrelogram and rotated versions of itself at 
\(30^\circ\), \(60^\circ\), \(90^\circ\), \(120^\circ\), and \(150^\circ\) are computed. 
The {\em gridness score} is then calculated as the difference between the 
lowest correlation value at \(60^\circ\) and \(120^\circ\) and the highest 
correlation value at \(30^\circ\), \(90^\circ\), and \(150^\circ\). 
Figure~\ref{fig:gridness} illustrates the underlying motivation for this 
measure. The two graphs show the results for successive correlations between the
masked autocorrelogram of a grid cell with rotated versions of itself in 
\(6^\circ\) steps. The graph shown on the right of figure~\ref{fig:gridness} is
an example for a grid cell with a highly periodic, triangular firing pattern 
resulting in high correlation values at multiples of \(60^\circ\) and low 
correlation values in between. In contrast, the graph on the left is an example
for a grid cell with a less regular firing pattern resulting in a much weaker
difference between the expected correlation maxima at \(60^\circ\) and \(120^\circ\)
and the expected correlation minima at \(30^\circ\), \(90^\circ\), and \(150^\circ\).
Thus, the difference between the lowest of the expected correlation maxima and 
the highest of the expected correlation minima provides a suitable measure of a
grid cells triangular periodicity. Sargolini et al.\@~\cite{Sargolini2006} 
classify all cells with a gridness score greater zero as grid cells. Others, 
e.g., Wills et al.\@~\cite{Wills2010} use thresholds determined by the 95th 
percentile of a shuffled gridness score distribution. Typically, the shuffled 
distribution is obtained by randomly shifting the spike times of each cell by 
more than \(20\) seconds and less than trial duration minus \(20\) seconds 
before calculating the cell's gridness score. This way the correlation with the 
animal's position is broken while the temporal firing characteristics are 
preserved. In case of Wills et al.\@~\cite{Wills2010} the resulting threshold 
was \(0.27\). Hence, unlike Sargolini et al.\@~\cite{Sargolini2006} Wills et 
al.~\cite{Wills2010} would not classify the cell shown in 
figure~\ref{fig:gridness} on the left as grid cell.

In contrast to the measures described so far, the {\em phase} of a grid cell is
not an absolute measure. It describes the relative displacement between the 
firing grids of two co-located grid cells, i.e, grid cells with similar spacing,
orientation, and field size. The relative displacement is determined by 
calculating the cross-correlation between the firing rate maps of the particular
grid cells. Due to the cells' similarity in spacing, orientation, and field size
the resulting cross-correlogram looks similar to the autocorrelogram of a single
grid cell, with the main difference that the central peak is offset from the
cross-correlogram's center. This offset is the relative {\em phase} between the
two grid cells.

The previously described measures are well established and widely used across
the grid cell literature. In particular, the gridness score introduced by 
Sargolini et al.\@~\cite{Sargolini2006} is the primary measure to identify neurons
as grid cells. As a result, cells with a gridness score below the respective 
threshold used in a study are typically not included in the analysis. In a 
recent study Krupic et al.\@~\cite{Krupic2012} introduced a different approach to
characterize the firing patterns of cells in the MEA and adjacent parasubiculum 
(PaS). They calculated the Fourier power spectrum of each cell's firing rate map 
to identify the main plane waves that give rise to the firing pattern. To reduce
the effects of noise they further filtered the power spectrum by subtracting the
50th percentile value of the power spectrum generated from a spatially shuffled 
version of the data and setting negative values to zero. Furthermore, main peaks
in the vicinity of a higher peak were treated as local maxima and ignored in the 
subsequent analysis. To qualify as a cell with a spatially periodic firing 
pattern the maximal component of the Fourier power spectrum had to 
exceed 95\% of all components in the power spectrum of shuffled data.
Using this measure, it is possible to not only identify grid cells, i.e., cells
with hexagonal firing patterns, but also other cells that exhibit firing 
patterns with a different spatial periodicity.

\begin{figure}[t]
  \centering
  \includegraphics[width=1.0\textwidth]{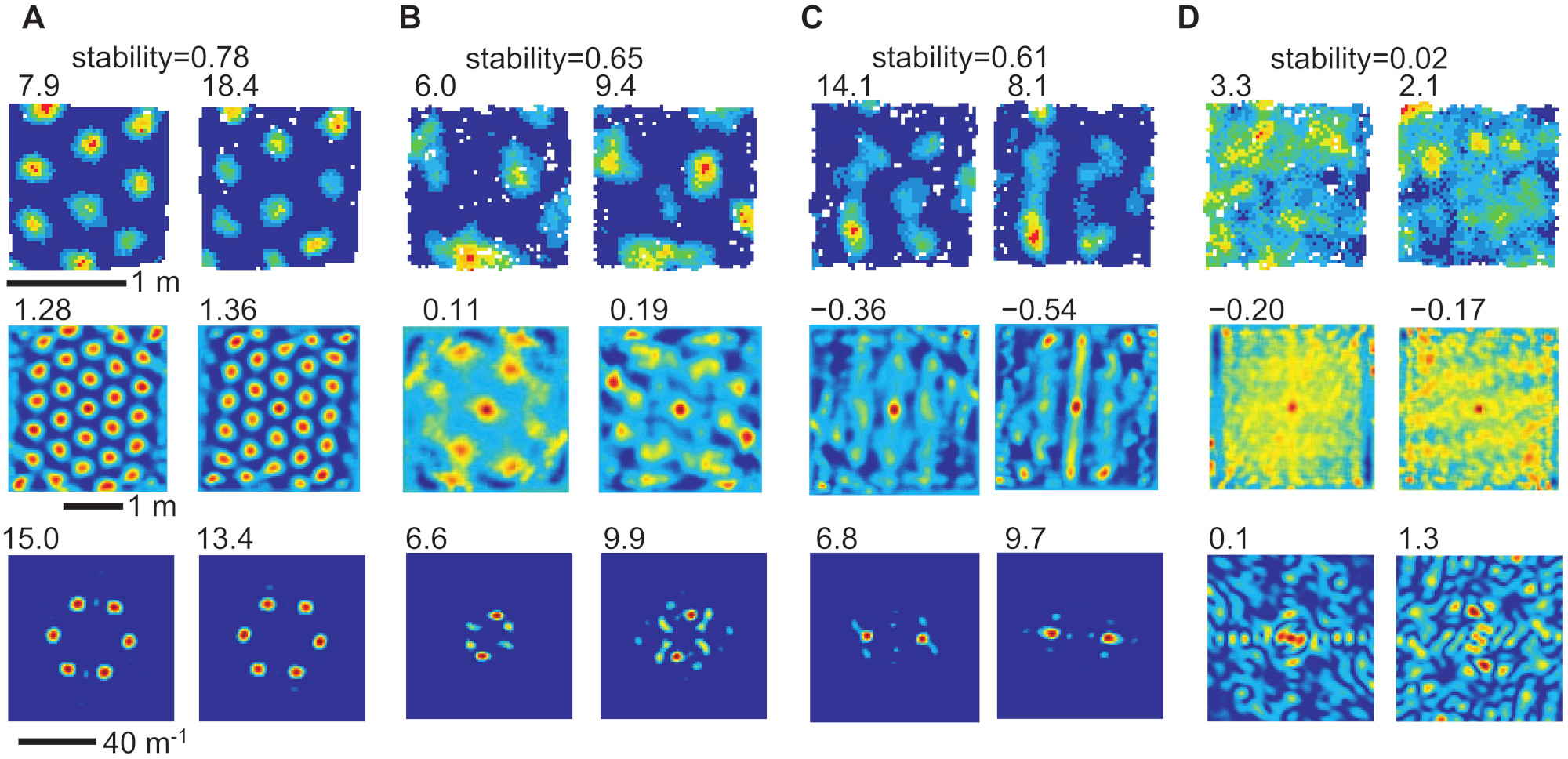}
  \caption[Spatial Periodicity]{Examples of three spatially periodic 
    ({\bf{A}-\bf{C}}) and one non-periodic cell ({\bf{D}}). Data for two 
    successive trials per cell are shown: unsmoothed firing rate maps (first 
    row), autocorrelograms of the firing rate maps (middle row), and filtered 
    Fourier power spectrograms (last row). Color scale from blue (low values) to
    red (high values). Stability of the firing patterns across trials is given 
    as Pearson product-moment correlation coefficient between the firing rate 
    maps. Peak firing rate, gridness score, and maximum Fourier power are 
    indicated above the particular maps. Figure adapted from Krupic et 
    al.\@~\cite{Krupic2012}.}
  \label{fig:grid_fourier}
\end{figure}

Krupic et al.\@ recorded 351 cells in seven rats from layers II and III of the MEA
and adjacent PaS. 65\% of all recorded cells in MEA and 79\% of all recorded 
cells in PaS were classified as spatially periodic cells. Of these 48\% in MEA 
and 18\% in PaS were classified as grid cells based on their gridness score. 
As expected grid cells exhibited three main components in the Fourier power 
spectrum which were separated by multiples of 60°. The other, non-grid cells had 
one to four main components with varying relative orientations and wavelengths. 
Figure~\ref{fig:grid_fourier} shows examples of firing rate maps, 
autocorrelograms, and Fourier power spectra for cells with periodic and 
non-periodic firing patterns. 

The main Fourier components of all spatially periodic cells were clustered 
around a limited number of orientations and wavelengths confirming the findings 
of Stensola et al.\@~\cite{Stensola2012} on the topographical organization of 
grid cells in MEA (see section~\ref{sec:toporg}). In addition, the use of the 
Fourier spectrum revealed that the orientations of the main Fourier components 
of spatially periodic non-grid cells were similar to the orientations of grid 
cells as well and only differed in a wider distribution of relative orientations
within the non-grid cells. 

The firing patterns of all cells classified as spatially periodic were more
stable than chance within as well as across trial days. Of these patterns the 
firing patterns of grid cells were the most stable. However, the number and 
relative orientation of the main Fourier components of some cells changed 
gradually over time such that, e.g., cells classified as grid cells became 
non-grid cells or vice versa. This change could be observed for trials within 
the same environment (11\% of spatially periodic cells changed) as well as 
across different\footnote{In this case ``different'' refers to a change from a
square to a circular environment.} environments (32\% of spatially periodic 
cells changed). Diehl et al.\@~\cite{Diehl2018} investigated the stability of
the spatial and non-spatial firing patterns of cells in the MEA in more detail
and could confirm the observations of Krupic et al. In particular, they observed
that between 9\% and 24\% of cells in MEA depending on cell classification and 
environment changed their spatial firing patterns and/or firing rate across 
recording sessions with inter-session time intervals ranging from 5 minutes to 
6 hours. In addition, they were able to characterize the change in firing 
patterns of individual cells as varying around a stable set point rather than 
random drift.

These results suggest, that grid cells may be just one instance in a continuum 
of spatially periodic cells~\cite{Diehl2017,Diehl2018}. This hypothesis could, 
e.g., provide an explanation for the multi-peaked, irregular firing patterns of 
early, putative grid cells in developing rats observed by Wills et 
al.\@~\cite{Wills2010} and Langston et al.~\cite{Langston2010} (see next 
section).

\section{Development}
\label{sec:development}

The development of spatial representations as expressed by head direction cells, 
place cells, and grid cells in the PHR-HF of young rats was recently 
investigated by Wills et al.\@~\cite{Wills2010} and Langston et 
al.\@~\cite{Langston2010}. Both teams used comparable experimental procedures 
and obtained consistent results. However, the teams differ slightly in their 
respective interpretations of the results.

Wills et al.\@~\cite{Wills2010} recorded putative place cells from CA1, as well as
putative grid and head direction cells from MEA in rats between the ages of P16
(postnatal day 16) and P30. During recording the rats foraged for food in a 
\(62cm \times 62cm\) box. Head direction cells exhibited strong directional tuning
and were found in adult-like proportion right from age P16, i.e., during the 
rat's first exploration of its environment\footnote{The rat's eyelids unseal at 
around P14 to P15.}. Similarly, place cells could also be observed in a 
significant proportion from day P16 with adult-like stability and quality in 
their firing pattern. Throughout the observed development period the number of
place cells increased steadily towards adult-like levels. In case of grid cells,
putative grid cells with multi-peaked firing fields could also be observed from
day P16. However, the recorded firing patterns were irregular at first. 
Significant proportions of cells with adult-like, hexagonal firing patterns 
emerged around P20 and increased fast to near-adult proportions by P22. Based
on these results Wills et al.\@ question the hypothesis that MEA grid cells 
provide the only spatial input to place cells. They point out that the observed
differential developmental time course suggest that the interconnectivity 
between grid cells and place cells develops only after the pups begin to explore
their environment. 

\begin{figure}[t]
  \centering
  \includegraphics[width=1.0\textwidth]{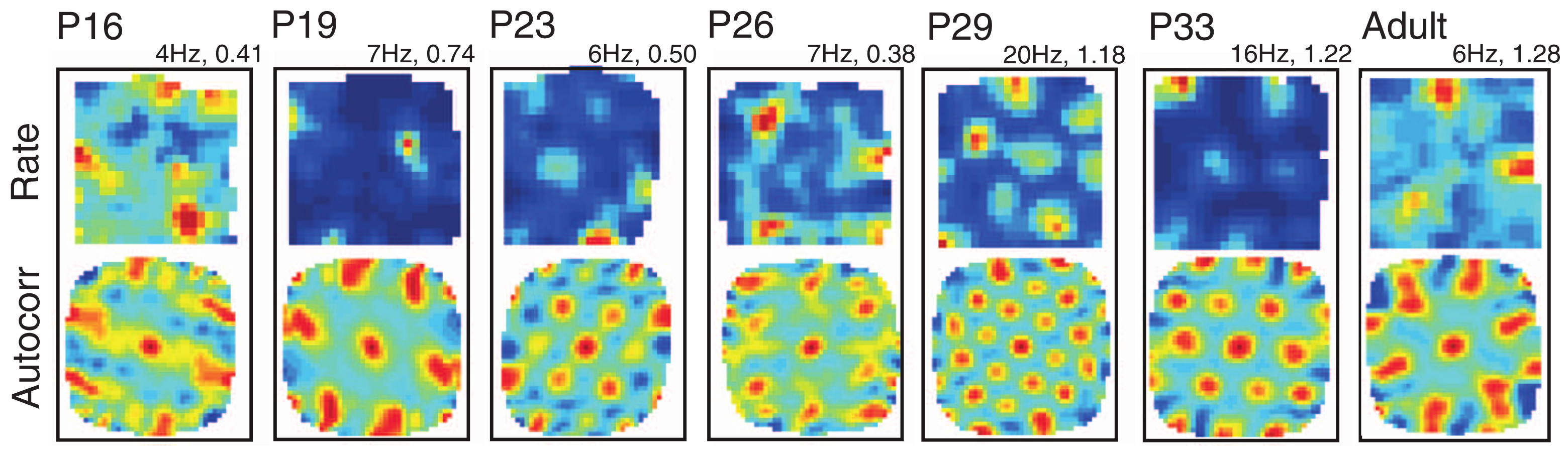}
  \caption[Grid Development]{Firing rate maps and their autocorrelations of
    grid cells in rats of ages between P16 and P33. Color scale from blue (low
    values) to red (high values). Peak firing rates and gridness scores 
    indicated above each rate map.
    Figure from Langston et al.\@~\cite{Langston2010}.}
  \label{fig:grid_dev}
\end{figure}

Langston et al.\@~\cite{Langston2010} recorded putative place cells from CA1,
putative grid cells from MEA, as well as putative head direction cells from
pre- and parasubiculum in rats between the ages P16 and P35. In addition they
recorded cells from an adult control group. During recording the rats foraged 
for food in a \(50cm \times 50cm\) box. In essence, their observations are 
consistent with the results obtained by Wills et al.\@~\cite{Wills2010}. 61.9\% of
all cells recorded in the pre- and parasubiculum in rats of age P15 and P16 
could be classified as head direction cells showing strong directional tuning. 
This proportion of HD cells was similar to the proportion found in the adult
control group (64.0\%). 41.4\% of the recorded cells in CA1 could be classified
as place cells in rats of age P16 to P18. This proportion continued to increase
steadily throughout development to adult-like levels at around 60\%. During 
development the stability of the firing fields generally increased within as
well as between trials. This finding matches earlier reports on the development
of place cells by Martin et al.\@~\cite{Martin2002}. In the MEA putative grid 
cells recorded in rats of age P16 to P18 showed multi-peaked, irregular firing 
fields. Yet, 12.8\% of the cells recorded at age P16 to P18 could already be 
classified as grid cells. The proportion of cells passing the grid cell 
criterion increased slightly during development to about 17.5\% in rats of age 
P31 to P34 being substantially lower than the proportion of grid cells measured 
in the adult control group (\(\sim\)~30\%). However, the periodic properties of 
cells classified as grid cells increased noticeably during development reaching 
gridness scores of near-adult levels at age P34. Figure~\ref{fig:grid_dev} 
illustrates this development showing the firing rate maps and their 
autocorrelograms of several grid cells in rats of increasing age. In contrast to
Wills et al.\@~\cite{Wills2010} Langston et al.\@ suggest that the three observed
cell types (HD cells, place cells, and grid cells) may interact from the outset
of exploration. They hypothesize that the observed, rudimentary grid cells 
provide a sufficiently patterned input to CA1 cells to enable the generation of
place-specific firing fields in the hippocampus.

\section{Topographical Organization}
\label{sec:toporg}

Grid cells in the MEA are topographically organized. Neighboring grid cells 
exhibit similar grid spacing, field size and grid orientation, but have 
dissimilar phases. Starting from the postrhinal border grid spacing and field 
size increase along the dorsoventral axis of the MEA. A similar, systematic 
change of grid orientation along this axis could not be 
observed~\cite{Hafting2005}.

\begin{table}[t]
  \centering
  \bgroup
	\def\arraystretch{1.33}
  \small
  \begin{tabular}{l  r@{.}l @{\;}r@{\;} r@{.}l 
                     r@{.}l @{\;}r@{\;} r@{.}l
                     r@{.}l @{\;}r@{\;} r@{.}l }  
    \multicolumn{1}{c}{} & \multicolumn{5}{c}{dorsal} & 
      \multicolumn{5}{c}{intermediate} & 
      \multicolumn{5}{c}{ventral} \\[0.3em]
    \hline 
    mean number of fields & 8 & 4 & \(\pm\) & 0 & 3 
                          & 5 & 9 & \(\pm\) & 0 & 4 
                          & 4 & 6 & \(\pm\) & 0 & 4 \\
    mean field width & 56 & 0 & \(\pm\) & 1 & 0 cm 
                    & 92 & 0 & \(\pm\) & 6 & 0 cm 
                    & 119 & 0 & \(\pm\) & 7 & 0 cm \\
    largest field width & 90 & 0 & \(\pm\) & 4 & 0 cm 
                    & 129 & 0 & \(\pm\) & 8 & 0 cm 
                    & 190 & 0 & \(\pm\) & 13 & 0 cm \\[0.75em]
    \hline
    minimum spacing & 91 & 0 & \(\pm\) & 12 & 0 cm 
                    & 202 & 0 & \(\pm\) & 24 & 0 cm 
                    & 269 & 0 & \(\pm\) & 47 & 0 cm \\
    median spacing & 171 & 0 & \(\pm\) & 13 & 0 cm 
                   & 301 & 0 & \(\pm\) & 23 & 0 cm 
                   & 370 & 0 & \(\pm\) & 46 & 0 cm \\[0.75em]
    \hline
    mean firing rate & 3 & 6 & \(\pm\) & 0 & 2 Hz 
                     & 4 & 5 & \(\pm\) & 0 & 4 Hz 
                     & 2 & 0 & \(\pm\) & 0 & 2 Hz \\
    peak firing rate & 21 & 3 & \(\pm\) & 0 & 9 Hz 
                     & 17 & 4 & \(\pm\) & 1 & 1 Hz 
                     & 11 & 4 & \(\pm\) & 0 & 9 Hz \\[0.75em]
    \hline
  \end{tabular}
  \normalsize
  \egroup
  \caption[grid scales and sizes]{
  	Summary of grid cell properties obtained by Brun et al.\@~\cite{Brun2008} for
    143 grid cells in 15 rats. Cells were partitioned into dorsal, intermediate,
    and ventral groups according to their position along the dorsoventral axis 
    of the MEA. 
  }
  \label{tbl:grid_scales}
\end{table}

In order to determine the range of grid spacings and field sizes present in the
MEA Brun et al.\@~\cite{Brun2008} recorded 143 grid cells in 15 rats that shuttled
back and forth on a 18m long, linear track. The cells were sampled from all 
entorhinal cell layers with an emphasis on the superficial layers (layer~II: 26,
layer~III: 35, layer~II or III: 40, layers V and VI: 42) and their locations
were distributed between \(1\%\) and \(75\%\) along the dorsoventral axis. To 
compare grid spacing and field size with respect to the cells' position the 
cells were grouped into a dorsal (\(0\%\) - \(25\%\), 55 cells), intermediate 
(\(25\%\) - \(50\%\), 59 cells), and ventral group (\(50\%\) - \(75\%\), 29 cells). 
Table~\ref{tbl:grid_scales} summarizes the obtained results. Grid spacing as 
well as field size increase from dorsal to ventral positions and, 
correspondingly, the number of firing fields along the 18m track decreases. In 
addition, an increase in field size appears to be accompanied by a reduction in 
peak firing rate. Grid spacing is characterized not only by median values but 
also by minimum values. The minimum values were provided as the algorithm used 
to detect the firing fields missed a substantial number of visually discernable 
fields due to low firing rates. Thus, the median values may overestimate the 
true grid spacing. 

Using the same experimental setup as Brun et al.\@~\cite{Brun2008} Kjelstrup et 
al.~\cite{Kjelstrup2008} could show, that the increase in grid spacing and field
size along the dorsoventral axis of the MEA is reflected in the field sizes
of place cells in CA3 which receive input from the MEA. The width of the place 
fields ranged from 1.41m in the dorsal region up to 13.59m in the ventral region
of CA3.

In a recent study Stensola et al.\@~\cite{Stensola2012} examined the 
topographic organization of MEA grid cells in more detail. In particular, they
investigated whether the increase in grid spacing and field size along the 
dorsoventral axis is continuous or modular, the latter option being indicated by
earlier experiments~\cite{Witter2006,Barry2007} and theoretical 
considerations~\cite{McNaughton2006,Fuhs2006}. In total 968 grid cells from 15 animals were
recorded while the animals foraged in 100cm to 220cm wide, square boxes. The 
high number of recorded cells per animal (up to 186 grid cells) and the use of
two sampling strategies that covered large parts of the MEA were key to enable
the determination whether the topographic organization of grid cells is 
continuous or modular. In every single animal a modular organization could be 
observed. Grid cells within a module share similar grid spacing, field size, and
grid orientation. Modules with increasing grid spacing and field size along the
dorsoventral axis overlap in their extent, i.e., the positions of grid cells
belonging to different modules are not separated and may interleave. Across all 
animals the distribution of mean grid spacing values covers a wide range with
no apparent peaks. However, within animals the scale relation between grid 
spacings of successive modules is governed by a fixed factor of approximately 
1.42 (\(\sqrt{2}\)) leading to a doubling of the area covered by each grid hexagon
between modules. The circumstance that the same scale ratio was found in all 
animals despite different, absolute values for the grid spacings implies that a
genetic mechanism is responsible for the scale relation while the different 
absolute values may be influenced or determined by external factors. 

Many of the hexagonal grid patterns observed by Stensola et al.\@ were elongated 
in one direction and it could be shown that grid cells sharing the same 
distortion also shared grid spacing, field size, and orientation, i.e., belonged
to the same module. By exploiting an experimental paradigm that provokes a {\em
rescaling} of the hexagonal pattern in grid cells (see 
section~\ref{sec:scaling}), it could also be shown that grid cell modules are
functionally independent. Each grid cell module exhibited the induced rescaling 
phenomena independent from each other, i.e., some modules exhibited rescaling
while others did not. This indicates that inputs based on the same environment
are processed independently by each module.

Stensola et al.\@ found a maximum of five grid scale modules per animal and they 
estimate that the total number of grid cell modules in the MEA is in the upper
single digit range.

\section{Functional Properties}
\label{sec:funcsec}

\subsection{Phase Precession}
\label{sec:phaseprec}

The phenomenon of {\em phase precession} refers to a specific relation between 
the firing behavior of an individual neuron and the overall, extracellular
activity, the electroencephalogram (EEG), of a brain region. In case of the rat
hippocampus two major classes of overall, extracellular activity can be 
observed. Behaviors like walking, running, swimming, rearing, or jumping are
accompanied by a characteristic, sinusoidal 7Hz to 12Hz activity called {\em 
theta activity} or just {\em theta} for short. Other behaviors like eating,
drinking, or grooming, i.e., behaviors that do not change the location of the 
animal, correlate with large irregular activity covering a broad spectrum of 
frequencies. During theta activity place cells fire in bursts at specific points
of the theta phase, e.g., at the trough. When an animal crosses the firing field
of a place cell the point of the theta phase at which the cell fires shifts with
the animal`s relative position within the place field. This shift in the timing
of the place cell's activity relative to the overall theta activity is called 
{\em phase precession}~\cite{OKeefe1993}.

Hafting et al.\@~\cite{Hafting2008} investigated whether entorhinal grid cells
exhibit phase precession as well. They recorded 174 MEA grid cells from 23 rats
while the rats ran back and forth on two linear tracks (235cm and 320cm, both 
10cm wide). The grid cells were identified in a two-dimensional open field 
environment in which grid spacings ranged from 30cm to 70cm.

Grid cells in MEA layer~II exhibited clear phase precession. On average grid 
cells started to fire at \(222 \pm 62^{\circ}\) when the animal entered a grid
field and stopped firing at \(59 \pm 78^{\circ}\) on exit labeling the peak of 
the theta phase with \(0^{\circ}\) and the trough with \(180^{\circ}\). The mean
slope of a best fit linear regression line on the individual firing events of
each cell was \(-2.77 \pm 0.31^{\circ}\,\mathrm{cm}^{-1}\). In layer~III of MEA 
25\% of all recorded grid cells showed phase precession over the complete theta 
cycle, another 25\% showed phase precession limited to the trough of the 
theta-phase, and 50\% exhibited no phase precession at all. On average layer~III
grid cells started to fire at \(195 \pm 81^{\circ}\) and stopped firing at \(192 
\pm 123^{\circ}\) of the theta-phase. The mean slope of the linear regression 
line was \(-0.078 \pm 0.283^{\circ}\,\mathrm{cm}^{-1}\). 

In order to test whether the observed phase precession of grid cells is 
independent of signals from the hippocampus the firing phases of grid cells were
recorded after inactivation of the hippocampus. This inactivation was achieved
by a local infusion of {\em muscimol}\footnote{5-aminomethyl-3-hydroxyisoxazole},
a \(\mathrm{GABA}_\mathrm{A}\) receptor agonist. After inactivation, the firing 
fields of layer~II grid cells remained spatially confined, albeit the firing 
fields of the grid cells became wider and less stable. Theta activity and grid 
cell phase precession was unaffected by the inactivation of the hippocampus as 
well. 

The question whether the phenomenon of phase precession is constitutive of grid
cell firing or not is controversial~\cite{Moser2014a}. A number of computational
models that classify as \emph{oscillatory interference} models use this property
to explain the hexagonal firing patterns of grid cells~\cite{Giocomo2011}. 
However, experimental evidence shows that grid cell firing in bats can be 
observed without any sign of concurrent theta activity contradicting the prior
of oscillatory interference models~\cite{Yartsev2011}. 

\subsection{Influence of External Cues}
\label{sec:extcues}

\begin{figure}[t]
  \centering
  \includegraphics[width=1.00\textwidth]{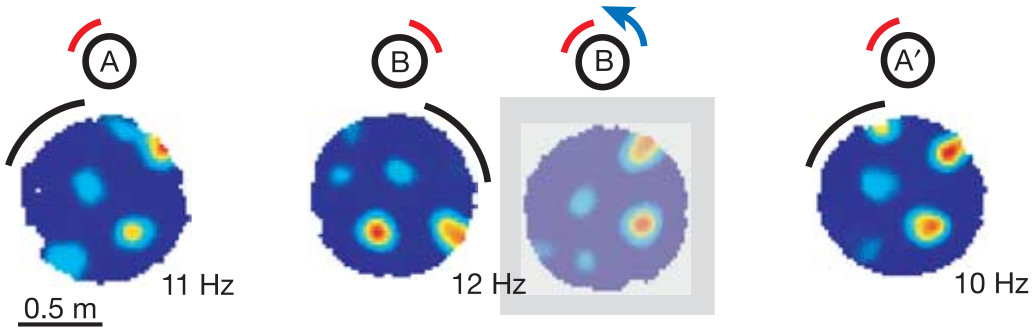}
  \caption[rotation]{Anchoring of the firing pattern to an external landmark. 
    Firing rate maps of a single grid cell recorded in a circular environment
    with a prominent visual cue (arc) attached to the environment's wall. Rate 
    maps are shown before rotation (A), after rotation by 90 degrees (B), and 
    after counter-rotation by 90 degrees (A'). Peak firing rates are given next 
    to the rate maps. The shaded rate map depicts the rate map of B artificially 
    counter-rotated by 90 degrees to ease visual comparison with A and A'.
    Figure adapted from Hafting et al.\@~\cite{Hafting2005}.}
  \label{fig:rotation}
\end{figure}

Hafting et al.\@~\cite{Hafting2005} were the first to characterize basic 
functional properties of grid cells. One of the key questions they address in 
this work is whether the locations of grid cell firing fields are determined by 
external landmarks (allothetic cues) or internal information about the rat's 
movement (idiothetic cues). They found that the firing fields of grid cells
stay constant across successive exposures to a single environment indicating a
strong influence of allothetic cues. To further investigate a possible anchoring
of the firing fields to external landmarks 24 grid cells in the dorsocaudal 
region of the medial entorhinal cortex (dMEC) in three rats were recorded. While
the rats chased food pellets in a circular environment, a prominent visual cue
at the wall of the environment was rotated by 90 degrees. The grid orientation 
of all 24 grid cells followed this rotation and only returned to their original
alignment when the visual cue was counter-rotated to its original position 
(fig.~\ref{fig:rotation}). 

Despite this strong influence of allothetic cues on the orientation and phase
of the grid cell firing fields external cues are not necessary to maintain the 
firing pattern itself. Hafting et al.\@ demonstrated this property by recording 
the activity of grid cells (4 rats, 33 cells) in total darkness (30 minutes)
preceded by a period of regular illumination (10 minutes). Besides a weak 
dispersal of the firing fields the overall grid pattern stayed intact during
the period without illumination. Hafting et al.\@ conclude that these findings
indicate that the grid pattern itself may result to a large extent from 
hardwired network mechanisms, but that the particular alignment of the pattern 
is determined by external landmarks. Kraus et al.\@~\cite{Kraus2015} 
investigated the maintenance of the grid pattern based on idiothetic cues and
elapsed time further. They recorded grid cells while rats were running on a 
motorized tread mill and could observe that grid cells were able to integrate
elapsed time as well as the distance that the animals had run.

\subsection{Realignment}
\label{sec:realignment}

A similar influence of external landmarks as described in the previous section
was already known to apply to hippocampal place cells~\cite{OKeefe1976}, i.e., 
a rotation of visual cues results in a corresponding rotation of a place cell's 
firing field. Likewise, the firing pattern of a place cell remains intact 
during locomotion in darkness. Further investigation of the influence of 
distant visual cues on place cell activity revealed the phenomenon of {\em 
remapping}~\cite{OKeefe1978a,Muller1987}. For a given environment a certain set
of place cells represents this particular environment. When the environment 
changes, e.g., when a rat is placed from one experimental environment into 
another, a different set of place cells gets recruited to represent the new 
environment. This switch from one set of place cells to another is termed {\em 
global remapping}~\cite{Leutgeb2005a,Jezek2011} and it only occurs if the
environment changes significantly. If, in contrast, the environment changes only
slightly, the same set of place cells is used but the maximum intensity with 
which individual place cells are active changes. This change of maximum activity
is termed {\em rate remapping}~\cite{Leutgeb2005a}.

Given the similar behavior of grid cells and place cells with respect to the 
influence of external landmarks, Fyhn et al.~\cite{Fyhn2007} investigated how
grid cells would behave during environment changes that reliably induce either 
{\em global remapping} or {\em rate remapping} in hippocampal place cells. 
Global remapping was induced by three protocols: switching between a square and 
a circular environment at a fixed location in one room, switching between 
similar square environments in two rooms with different background cues, and 
switching between light and darkness in a single, square environment. Rate 
remapping was induced by a single protocol in which the colors of the walls of a
single, square environment were changed. Neuronal activity was recorded in 
dorsocaudal MEA and/or dorsal CA3 in 19 rats. The protocols reliably induced 
global and rate remapping in all trials in which CA3 place cells were recorded. 

\begin{figure}[t]
  \centering
  \includegraphics[width=1.00\textwidth]{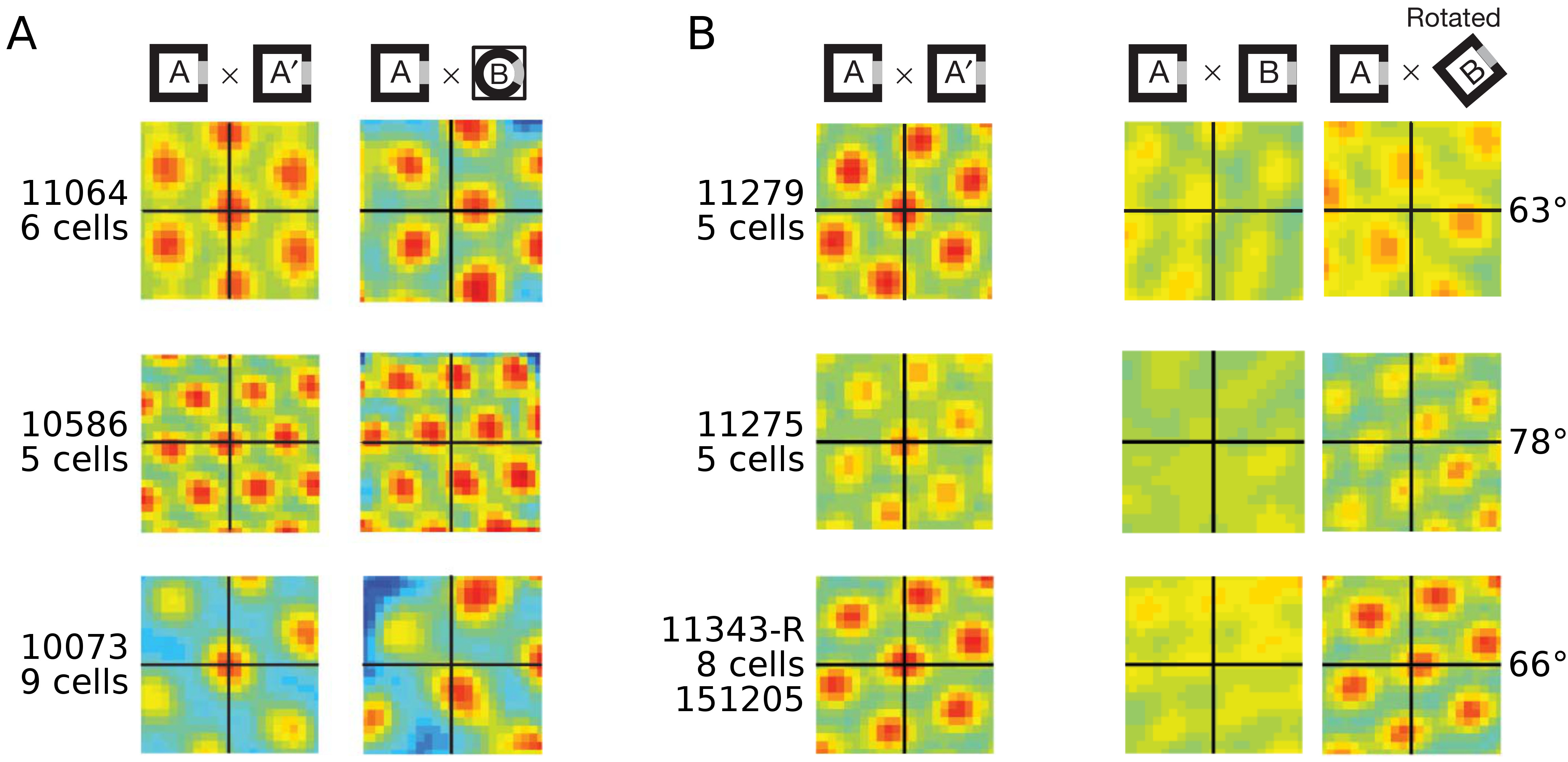}
  \caption[realignment]{Illustration of grid cell realignment. The maps show 
    cross-correlations of firing rate maps for small populations of grid cells
    (5 to 9 cells) in individual animals (5-digit numbers). {\bf{A}}: Switching
    between a square and circular environment causes a shift of the grid pattern
    while spacing, orientation, and relative phases are preserved. {\bf{B}}: 
    Switching between square environments located in different rooms causes not
    only a shift of the grid pattern but also a rotation of the pattern. The 
    angle of rotation was determined by successively rotating the firing rate
    maps of environment B until the resulting pattern in the cross-correlogram
    had maximal grid structure.
    Figure adapted from Fyhn et al.\@~\cite{Fyhn2007}.}
  \label{fig:realignment}
\end{figure}

Global remapping induced by the first protocol, i.e., alternating between
square and circular environments, led to an absolute shift of the grid up to
one-half grid spacing while the distributions of grid spacing, grid orientation,
and relative phases were preserved (fig.~\ref{fig:realignment}A). The directions
of the grid shifts were uniformly distributed across experiments. However, for 
individual cells that were recorded over several days the direction and 
magnitude of the grid shift remained constant. The second protocol, constituting
a stronger environment change, led not only to a shift of the grid pattern but 
also to a grid rotation (fig.~\ref{fig:realignment}B). Additionally, in some 
cases (three out of seven for the first protocol, five out of eight for the 
second protocol) the overall grid spacing slightly scaled between environments. 
These results indicate that grid cell ensembles do not change during global 
remapping and instead perform a {\em realignment} of grid orientation and 
position. They further suggest, that the relative phases of grid cells within a 
local ensemble may be in a rigid relationship (this hypothesis was recently
supported by results of Yoon et al.\@~\cite{Yoon2013}). The third global 
remapping protocol was used with two rats that had electrodes in MEA as well as 
CA3 to determine whether grid cell realignment and place cell remapping are 
coincident. In one animal global remapping of place cells coincided with an 
equally fast grid cell realignment, whereas in the second rat place cells and 
grid cells maintained their firing fields after a switch from the dark to the 
light condition. In the latter case, global remapping and realignment could be 
triggered by temporarily interrupting the movement of the rat by placing it for 
one minute on a pedestal. Fyhn et al.\@ argue that this delayed remapping 
reflects a continued influence of self-motion information on the location of 
place cell and grid cell firing fields.

Contrary to the global remapping condition no change of grid cell activity could
be observed during rate remapping. The results obtained by Fyhn et al.\@ suggest
two main implications. First, as the ensemble of grid cells and their relative, 
spatial relationship does not change across environments, similar paths of an 
animal in different environments are mapped onto similar sequences of grid cells
supporting the hypothesis that grid cells are part of an universal metric for 
path-integration-based navigation. Second, as hippocampal CA3 receives both 
direct and indirect input from the MEA the realignment of grid cells could serve
as a basis for the remapping occurring in hippocampal place 
cells~\cite{Monaco2011}. 

A number of recent publications provide further insight into the relations 
between MEC and hippocampus. Zhang et al.\@~\cite{Zhang2013} used a combined
optogenetic-electrophysiological strategy to map out the functional inputs to 
hippocampal place cells that originate in the entorhinal cortex. Their main 
finding suggests that hippocampal place cells receive input from a wide variety
of functional cells in the MEC including grid cells, border cells, head 
direction cells, irregular spatial cells, and non-spatial cells. Miao et 
al.\@~\cite{Miao2015} investigated the influence of partial MEC inactivation on
hippocampal place cells. They induced the partial inactivation by a 
pharmacogenetic approach and could observe hippocampal remapping in CA3 as 
response. The inactivation did not influence the shape or size of the observed
place fields, but their distribution at the ensemble level changed 
significantly. Importantly, this remapping took place almost instantaneously.
In contrast, Rueckemann et al.\@~\cite{Rueckemann2015} performed a similar
experiment, but observed a slow and lasting change in the place cell ensemble
after a partial inactivation of MEC. However, Rueckemann et al.\@ used an 
optogenetic instead of a pharmacogenetic strategy to partially silence the MEC
and, more importantly, they observed hippocampal place cells in CA1 instead of
CA3. These differences in the experimental setup may explain the diverging 
results regarding the time course of the observed remapping. Marozzi et 
al.\@~\cite{Marozzi2015} explored the effects of non-metric context changes, 
i.e., changes in color and smell of the environment on grid cell firing 
patterns and could observe a realignment of the grid phase as response. Finally,
Olafsdottir et al.\@~\cite{Olafsdottir2016} addressed the question if and how
the activity of grid cells is involved in the well known phenomenon of {\em 
hippocampal replay}. They recorded grid cells in layers V and VI of the MEC and
could show that these cells were spatially coherent with hippocampal place cells
during replay. The grid cells encoded the particular locations with a 11ms 
delay, indicating monosynaptic feedback connections from hippocampus to layers
V and VI of MEC.

\subsection{Rescaling}
\label{sec:scaling}

The slight scaling effects observed by Fyhn et al.\@~\cite{Fyhn2007} during 
realignment can have a much larger magnitude under certain circumstances. Barry 
et al.\@ discovered and investigated these stronger {\em rescaling} phenomena in 
two different contexts~\cite{Barry2007,Barry2012}. 

\begin{figure}[t]
  \centering
  \includegraphics[width=1.00\textwidth]{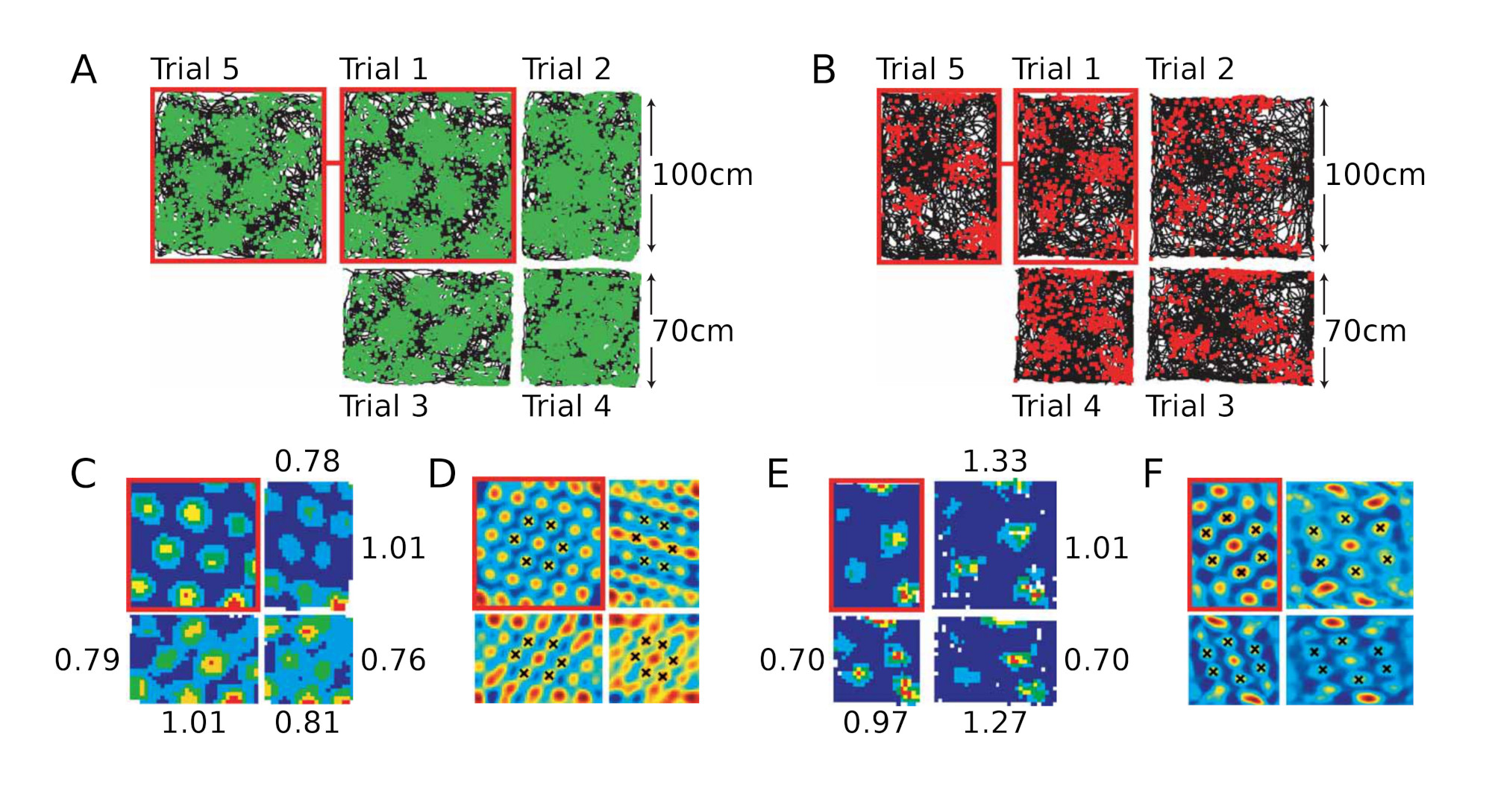}
  \caption[rescaling]{Illustration of grid cell rescaling. 
    {\bf{A,B}}: Firing events (green and red dots) of single grid cells 
      superimposed on the path (black lines) of two rats in scaled environments.
      Trials 1 and 5 took place in the environment to which the rats were 
      accustomed to (red outline).
    {\bf{C,E}}: Color-coded (blue: low, red: high) firing rate maps. Grid 
      scaling for each dimension indicated by numeric labels.
    {\bf{D,F}}: Color-coded autocorrelograms of the firing rate maps. Peaks of
      the central hexagons are marked with black crosses.
    Figure adapted from Barry et al.\@~\cite{Barry2007}.}
  \label{fig:rescaling}
\end{figure}

The first context in which a significant rescaling of grid cell firing 
patterns can be observed is the geometric deformation of a known 
environment~\cite{Barry2007}. Barry et al.\@ trained six rats to be accustomed to 
either a square (\(1m \times 1m\), 3~rats) or a rectangular (\(0.7m \times 1m\), 
3~rats) environment. Subsequently, the trained rats were exposed to scaled 
versions of these environments while the activity of MEA grid cells was 
recorded. The square (s) and rectangular (r) environments were scaled 
horizontally (\(s: 0.7m \times 1m \;\vert\; r: 1m \times 1m\)), vertically (\(s: 1m
\times 0.7m \;\vert\; r: 0.7m \times 0.7m\)), and in both directions (\(s: 0.7m 
\times 0.7m \;\vert\; r: 1m \times 0.7m\)). In all cases the recorded firing 
patterns of the grid cells showed strong rescaling in the direction in which the
particular environment was scaled, though with a lesser magnitude 
(fig.~\ref{fig:rescaling}). Re-exposure to the unscaled, familiar environment 
(Trial 5) showed, that the rescaling effect is not permanent. On average grid 
patterns rescaled by 47.9\% with respect to the change of the environment. In 
addition, cases of uni-directional scaling, i.e., in either horizontal or 
vertical direction, were accompanied by a small, opposite rescaling (7.9\% on 
average) of the grid patterns in a direction orthogonal to the change of the 
environment. 

Barry et al.\@ hypothesize that environmental features like boundaries become 
associated with the grid pattern over time, such that a sudden deformation of 
the environment causes a corresponding deformation of the grid pattern. However,
with an increasing number of trials, grid patterns in the changed environment 
rescaled with lower and lower magnitude indicating that rescaling is not solely 
a reflection of the environment's changed geometry but also a reflection of the 
animal's increasing familiarity with the modified environment. Barry et al.\@
suggest that this reduction in rescaling reflects a tendency of the grid cells 
to revert to an intrinsic grid scale. 

The second context in which rescaling of grid cell firing patterns was observed
is exposure to novel environments~\cite{Barry2012}. In the corresponding study
MEA grid cells of eight rats were recorded while the animals foraged in \(1m 
\times 1m\) environments. The rats underwent five trials (20 minutes each) per 
day for up to seven consecutive days. Each day, the rats were first exposed to a 
familiar\footnote{familiar := a minimum of 100 minutes of prior exposure} 
environment, followed by three novel environments, and finally the familiar 
environment again. Novel environments differed in texture, visual appearance, 
and odor. Exposure to the first novel environment on the first day caused 
realignment (rotation and shift) as well as rescaling of all recorded grid 
patterns. On average the grid patterns scaled up by 37.3\% (min.:~10.5\%; 
max.:~71.1\%). In addition, the average gridness score dropped from 0.65 in the
familiar environment to 0.04 in the novel environment as the hexagons of the 
grid patterns were less circular in the novel environment. Exposure to the 
second and third novel environment resulted in rescaling with less magnitude, 
i.e., grid patterns in the fourth trial scaled up 21.3\% on average. This trend
continued over subsequent days. On the second day the average increase was 
16.2\% and on day five no grid cell showed any discernable rescaling. In some 
cases the firing pattern of individual grid cells already showed no apparent 
rescaling after three days.

In order to examine if place cell remapping co-occurs with grid cell rescaling
another seven rats were implanted with electrodes recording simultaneously in
MEA and CA1. The rats were tested by a similar protocol as described before, yet
the environments used were not identical. Upon first exposure to a novel 
environment the firing patterns of the recorded grid cells scaled up by 33.9\% 
on average, and the average gridness score decreased from 0.96 to 0.31. 
Simultaneously recorded place cells showed an immediate and complete remapping
and the average size of place fields in the novel environment increased by 
28.8\%. Furthermore place fields in the novel environment appeared to be less 
stable and more fraying. A subsequent weakening of grid pattern rescaling in the
second and third novel environment could be observed as well (trial~3: 14.6\%; 
trial~4: 16.8\%) and was accompanied by a similar, though smaller decrease of 
place field scaling (trial~3: 11.3\%; trial~4: 7.6\%\footnote{This measurement 
did not reach a significance level of 0.05 (0.07).}).

In summary these results show that in novel environments the firing patterns of
grid cells as well as place cells expand and become less regular. With 
increasing exposure to the new environments the firing fields of grid cells and 
place cells reacquire the properties seen in familiar environments, i.e., they 
become more regular and smaller in spatial scale. As a further mechanism that 
may underlie these rescaling phenomena Barry et al.\@ propose a possible influence
of the neuromodulator acetylcholin (ACh) which is implicated in novelty 
detection. This hypothesis is supported by the fact that co-recorded grid cell
patterns scale up by similar amounts.

\subsection{Influence of Environmental Geometry}
\label{sec:environ}

The phenomena described in the previous sections were further investigated by
Krupic et al.\@~\cite{Krupic2015}. In contrast to previous reports they could 
show that environmental geometry can have a lasting influence on the firing 
patterns of grid cells. In a first experiment they demonstrate that the geometry
of the experimental arena can control grid orientation and override the 
influence of prominent distal cues if the geometry is {\em polarized} as, e.g.,
in the case of a square environment. If such an experimental enclosure is 
rotated \(45^\circ\), the grid orientation follows this rotation (mean grid 
rotation \(42.5^\circ \pm 2.9^\circ\)) despite the presence of stationary distal
cues. However, if the enclosure is rotated \(90^\circ\), the grid orientation 
remains unchanged (mean grid rotation \(1.1^\circ \pm 0.9^\circ\)). The latter 
case indicates that other local cues like smells or textures do not influence 
grid orientation. Further investigation (275 grid cells, 41 rats) showed that
grid orientation aligns to the walls of square enclosures at a mean angle of
\(8.8^\circ \pm 0.6^\circ\) while the results for unpolarized environments 
were less clustered. As one consequence of this alignment the relative 
orientation between different grid cell modules should be either \(0^\circ\) or 
\(30^\circ\) in square environments due to the \(60^\circ\) symmetry of the grid
pattern in relation to the \(90^\circ\) angles of the environment. Apart from a 
few intermediate values the observed relative orientations confirmed this 
hypothesis. Yet, these relative orientations remained stable in circular and 
hexagonal environments as well indicating that anatomically close grid cell 
modules can act coherently across different environments. The findings of Krupic
et al.\@ regarding the alignment of grid orientation to enclosure walls matches 
the findings of a similar study by Stensola et al.\@~\cite{Stensola2015}. The 
latter hypothesize that the alignment might be a way to minimize symmetry with 
the borders of the environment. 

Krupic et al.\@ also repeated the rescaling experiments of Barry et 
al.\@~\cite{Barry2007} described above, but they switched from a square to a 
trapezoid environment instead of a rectangular one. Their observations matched
those of Barry et al.\@ as firing fields expanded in response to the novel 
environment. Yet, contrary to the observations of Barry et al.\@ this expansion
reduced only slightly afterwards and firing fields remained 20\% to 30\% larger
compared with the firing field sizes in a concurrently recorded square 
environment. In addition, grid patterns in the trapezoidal enclosure showed a
permanent decrease in gridness score due to more elliptical, hexagonal patterns
and less evenly distributed firing fields. Based on these results Krupic et al.\@
conclude ``that most assumptions about the invariant nature of grid cell firing
are invalid'' and that ``in particular the role of environmental boundaries 
has been underestimated''.

\subsection{Fragmentation}
\label{sec:fragmentation}

\begin{figure}[t]
  \centering
  \includegraphics[width=0.75\textwidth]{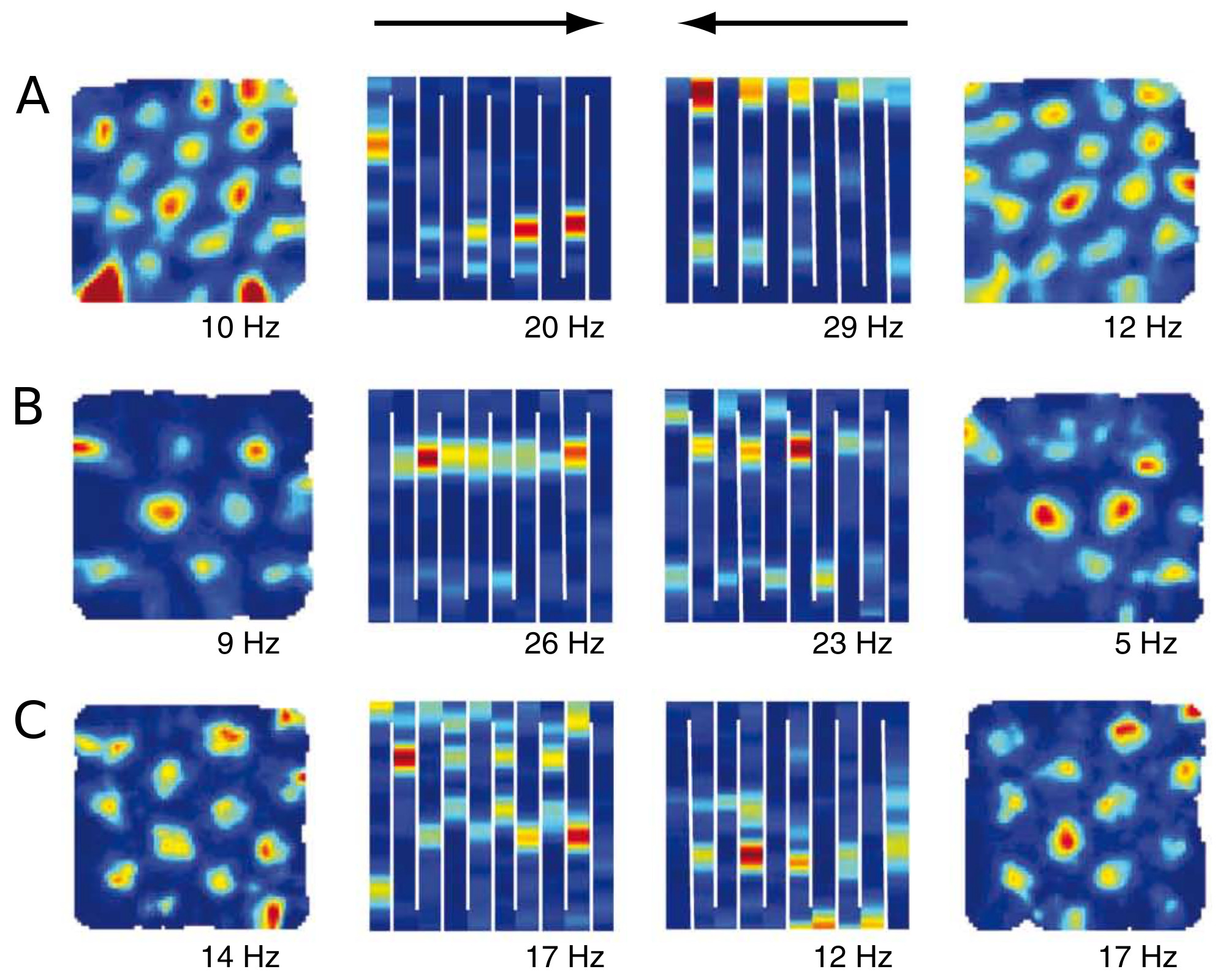}
  \caption[Grid Cell Fragmentation]{ Firing rate maps of three grid cells (rows
    A, B, C). Firing rates are represented by blue (low) to red (high) colors.
    Peak firing rates are given below the maps. The first and last columns show 
    the firing rate maps in the open-field, while the second and third columns 
    show the firing rate maps in the hairpin maze according to the animal's 
    running direction (indicated by black arrows).
    Figure adapted from Derdikman et al.\@~\cite{Derdikman2009}.}
  \label{fig:grid_frag_1}
\end{figure}

The phenomena of realignment and rescaling of grid cell firing patterns were 
observed and examined in rectangular or circular environments in which the rats 
could move around freely without encountering any obstacles. However, real 
environments are more likely to consist of multiple, connected subenvironments.
To investigate whether grid cells exhibit a continuous or fragmented firing 
pattern across such subdivided environments Derdikman et 
al.~\cite{Derdikman2009} recorded MEA grid cells of rats that ran through a 
hairpin maze consisting of several corridor-like passages connected by sharp,
\(180^\circ\) turns (white bars in fig.~\ref{fig:grid_frag_1}). The experimental
protocol consisted of four consecutive, twenty minute trials per day. In the 
first and last trial each rat foraged in an open-field 
box (\(1.5m \times 1.5m\)). In between, i.e., in the second and third trial, 
each rat ran back and forth in a hairpin maze which was inserted into the 
open-field box. A total of 105 MEA grid cells in 16 rats were recorded.

\begin{figure}[t]
  \centering
  \includegraphics[width=1.0\textwidth]{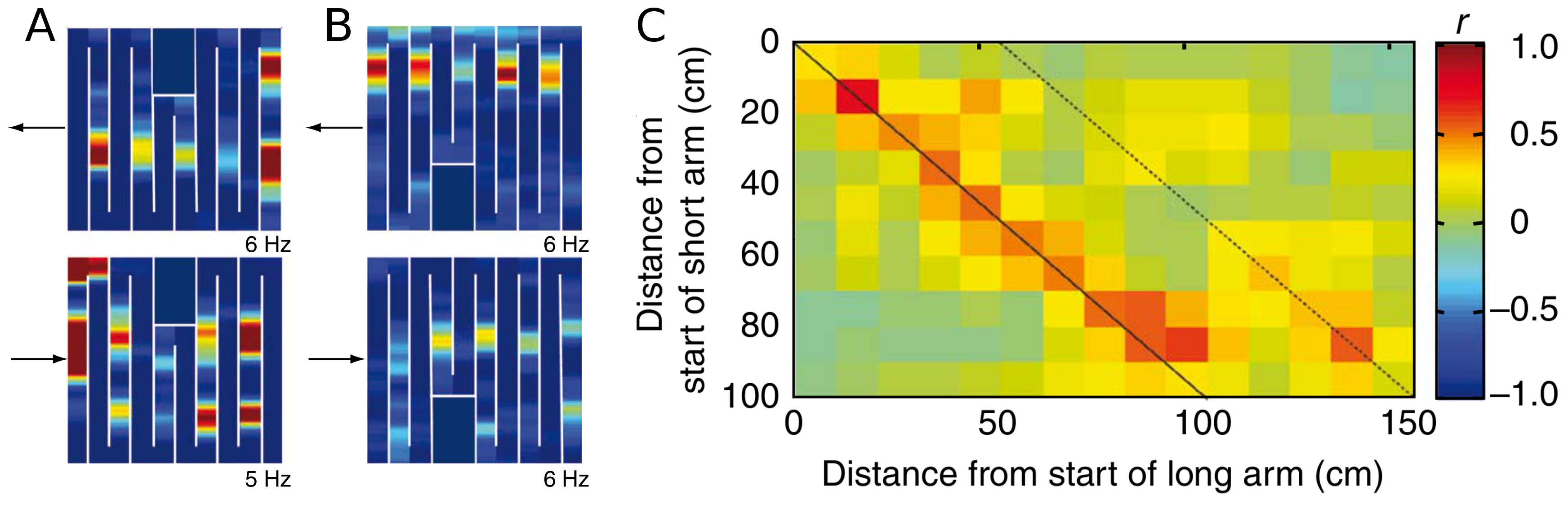}
  \caption[Reset Anchoring]{
    Modified hairpin maze with two arms shortened by \(50cm\).
    {\bf{A,B}}: Firing rate maps of two grid cells recorded in a modified 
    hairpin maze. Firing rates are represented by blue (low) to red (high) 
    colors. Peak firing rates are given below the maps. The animal's running 
    direction is indicated by black arrows.
    {\bf{C}}: Correlation between the activity (10cm bins) in a short arm with 
    the mean activity (10cm bins) in the corresponding long arms.
    Figure adapted from Derdikman et al.\@~\cite{Derdikman2009}.}
  \label{fig:grid_frag_2}
\end{figure}

Within this experimental setup grid cells showed the expected triangular firing 
pattern in the open-field box and lost this pattern in the hairpin environment
(fig.~\ref{fig:grid_frag_1}). In the hairpin environment the firing fields
of grid cells were located at similar positions in individual maze arms relative
to the rats running direction, i.e., firing fields in maze arms with equal 
running directions had similar positions while firing fields in maze arms with 
opposite running directions had differing positions in general. This result 
suggests that grid cell firing patterns reset at the turning points of the 
hairpin maze. To test whether the firing fields were anchored to the preceding or
upcoming turning point, two arms of the hairpin maze were shortened by \(50cm\)
(fig.~\ref{fig:grid_frag_2}). The resulting firing activity of each 
shortened arm was correlated with the activity of all corresponding long
arms using \(10cm\) wide bins and a progressive shift of the short arm to cover
any possible alignment. Figure~\ref{fig:grid_frag_2}C shows exemplarly the 
correlogram of one short arm with the mean activity of all corresponding long 
arms. In general, the short arm correlates more strongly with the longer arms
if it is aligned to the preceding wall, i.e., the activity pattern within a 
short arm is more similar to the start of a long arm than to its end with 
respect to the shared wall. However, the end of the short arm also correlates 
to some degree with the end of the long arms, i.e., when aligned to
the upcoming wall. Together, these results indicate that the location of the
firing fields is possibly determined by two mechanisms: in part by a form of 
path integration after a turning point, and in part by some form of 
environment-based alignment towards the upcoming turning point.

To exclude the possibility that the observed firing patterns could be caused by
the running pattern of the animals, the rats were trained to run the path of the
hairpin maze in the open-field environment without the presence of the maze 
walls. The observed grid cell firing pattern matched the firing pattern during
the random foraging task in the open-field box. Thus, excluding a behavioral
cause for the fragmentation seen in the hairpin maze.

Derdikman et al.\@ examined also whether the observed resetting of the grid cell 
firing fields would influence hippocampal place fields. They recorded 111 place
cells in CA3 (4 rats) and 47 place cells in CA1 (3 rats) and recorded their 
activity using the same experimental protocol as above. Place cells recorded 
while the animal was in the hairpin environment showed a fragmentation of their 
firing fields similar to that of grid cells. Place fields in arms with the same
running direction were highly correlated, whereas place fields in arms with 
different running direction were only weakly correlated. This co-occurring
fragmentation of MEA grid cell firing fields and CA1/CA3 place fields is another
indication for a possible coupling between the spatial representations in 
the MEA and the hippocampus.

\begin{figure}[!t]
    \begin{center}
        \subfigure[]{\includegraphics[width=0.5\textwidth]{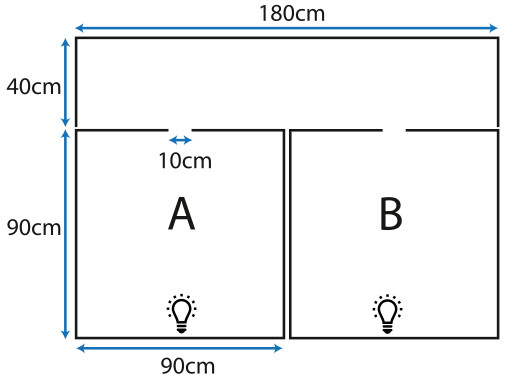}}
        \subfigure[]{\includegraphics[width=1.0\textwidth]{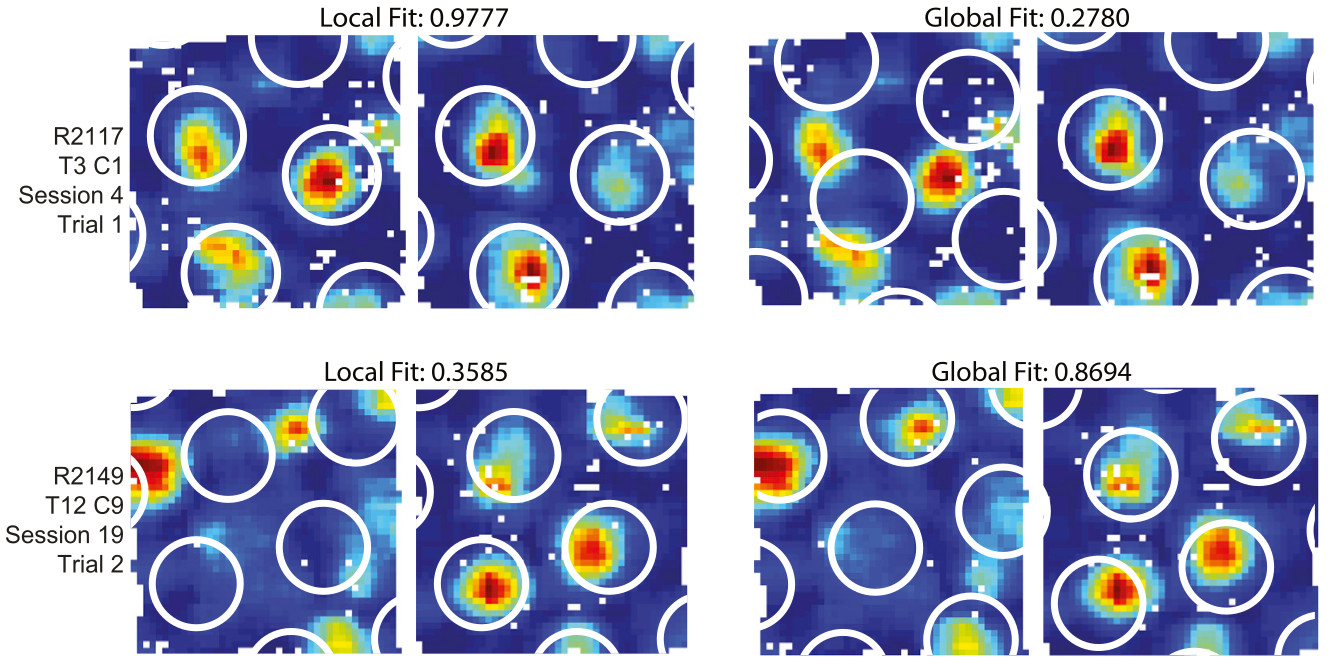}}
        \caption[Global Representation]{Prolonged exposure to an environment can
          lead to a global representation in grid cells. ({\bf{a}})~Schematic of
          the experimental setup. Two perceptually identical environments A and 
          B were connected by a corridor. ({\bf{b}})~Example of two firing rate 
          maps (rows) fitted to a local and a global model (columns). The rate
          map from an early session (top row) was better explained by the local
          model, whereas the rate map from a late session (bottom row) was 
          better explained by the global model.
          Figures adapted from Carpenter et al.\@~\cite{Carpenter2015}.}
        \label{fig:globref}
    \end{center}
\end{figure}

The observations made by Derdikman et al.\@~\cite{Derdikman2009} and Krupic et 
al.\@~\cite{Krupic2015} suggest, that contrary to prior assumptions grid cell
firing may primarily be determined by local environmental cues instead of 
forming a single, global representation of the environment. To test this 
hypothesis Carpenter et al.\@~\cite{Carpenter2015} devised an experiment where
rats foraged in two perceptually identical enclosures connected by a corridor 
(fig.~\ref{fig:globref}a). The walls of the environment were painted black and
distal cues were hidden by black curtains. Lighting was provided by two single
lights on the south wall of each enclosure. To control for other sensory cues 
such as odor, which could help the animal to distinguish between the 
compartments, the floor was rotated, the compartments swapped, and the 
environment cleaned in the middle of each recording session, i.e., each session
consisted of a 40 minute trial (Trial 1), then the reconfiguration and cleaning 
of the environment, and a second 40 minute trial (Trial 2).

To determine if the firing pattern of a grid cell would locally replicate in 
each compartment or if it would globally cover both compartments, Carpenter et 
al.\@ fitted a local and a global model to the firing rate map. In case of the
local model the fitted grid had the same orientation, scale, and phase in both
compartments. In case of the global model the fitted grid covered both 
compartments, i.e., the phase remained continuous across compartments. During
the first recording sessions, the firing rate maps of the recorded grid cells 
(85 cells, 8 rats) were best described by the local model confirming the 
observations of Derdikman et al.\@ and Krupic et al.\@ described above 
(top row in fig.~\ref{fig:globref}b). However, with prolonged exposure to the
environment, the fit of the local model steadily decreased while the fit of the
global model increased, i.e., the observed grid cells formed a global instead of
a local representation of the environment with experience (bottom row in 
fig.~\ref{fig:globref}b). The transition from a local to a global representation
appeared to occur gradually and continuously. In particular, Carpenter et al.\@
were able to track the adjustment of the firing fields of a single grid cell
over a prolonged period of time observing that the firing fields ``appeared to
shift continuously, rather than undergo a sudden transformation''. 

Wernle et al.\@~\cite{Wernle2017} conducted a similar experiment in which rats
were first trained to forage for food in two adjacent environments A and B that
shared common distal cues and were separated by a central wall. After training 
the rats were then exposed to the joint environment AB by removing the 
separating wall. Wernle et al.\@ recorded 128 grid cells (10 rats) that were
active in all three environments (A, B, and AB). In agreement with the 
observations of Carpenter et al.\@ the majority of the grid cells represented
the environments A and B with separate grid patterns that differed in phase, and 
occasionally also in orientation and scale. After removal of the central wall 
the separate grid patterns merged into a single, locally coherent grid pattern.
In contrast to the observations of Carpenter et al.\@ this reorganization of the
grid pattern happened on a fast time scale, i.e., within minutes during the
first trial in the joint environment. Furthermore, Wernle et al.\@ report that
firing fields near the outer borders of the joint environment remained anchored 
to their positions during reorganization while firing fields near the removed 
central wall moved away from their original locations in order to self-organize
into a locally coherent grid pattern.

In summary, these results indicate that the early hypothesis of {\em stable} 
grid patterns that span the entire environment and serve as a metric for 
space~\cite{Moser2008a,Moser2008b} has to be revised to reflect these more 
recent observations of dynamic changes in grid cell firing patterns and their 
anchoring to environmental cues. Carpenter and Barry~\cite{Carpenter2016}
provide a first discussion in this regard.

\section{Neuronal Structure}
\label{sec:neurostruct}

Neurons with grid-like firing patterns were found in all populated layers of the
MEA~\cite{Hafting2005,Sargolini2006} as well as the PrS and 
PaS~\cite{Boccara2010} (see Tang et al.\@~\cite{Tang2015} for a recent
contradictory report on layer III). Interestingly, the subsequently described 
morphology of neurons found in these layers is not homogeneous, i.e., grid-like 
firing patterns are exhibited by cells with varying 
morphologies~\cite{Rowland2018,Gu2018}. It is an open question whether grid 
patterns are generated independently by various types of cells or by just a 
single cell type which then projects a grid signal to other cells that merely 
forward the signal. 

\subsection{Neuron Morphology in the Entorhinal Cortex}
\label{subsec:morphology}

The entorhinal cortex is the origin of a prominent pathway, the {\em perforant
path}, that connects the EC with all regions of the hippocampus. In addition,
the EC receives inputs from various regions of the brain including the 
neocortex~\cite{Witter2000a}. As a consequence, the EC is commonly thought of as
the gateway to the hippocampus. As such the neuronal structure of the EC has 
been extensively studied even before the discovery of grid cells. If not noted 
otherwise the neuronal structures reported below refer to the rat brain. As the 
PHR-HF is constantly present in all mammalian species with little variation 
during its phylogenetic development, insights into the neuronal structures of 
rat EC can be generalized to other mammalian species, e.g., humans to a certain 
degree~\cite{Insausti1993}. The following paragraphs will briefly summarize the 
key morphological properties of principal neurons found in the EC layer by layer
and provide an overview of other structural properties like local microcircuits 
and indications of a possible columnar organization of the EC.

The overall structure of the EC comprises six layers. Layers I to III are 
referred to as {\em superficial} layers whereas layers IV to VI are referred to 
as {\em deep} layers. Layer I contains only few neuron soma and is mainly 
occupied by dendritic and axonal processes originating from cells located in 
lower layers. Layer II is densely populated by stellate cells and small pyramidal 
cells as its principal neurons~\cite{Lingenhoehl1991}. In LEA layer~II splits 
into Layers IIa and IIb. Layer IIb is a continuation of MEA layer~II whereas 
layer~IIa is located more superficially and consists of stellate cells forming
local clusters or ``islands''~\cite{Koehler1986}. The principal cells of layer
III are mostly pyramidal cells. Layer III is relatively thick and is followed by
cell poor layer IV which contains only sparsely scattered pyramidal cells. The
relatively thick layer V hosts mostly small pyramidal 
neurons~\cite{Lingenhoehl1991}. Both layer IV and layer V are thicker in LEA 
than in MEA due to a higher neuron count and a less dense packaging of the 
cells in LEA~\cite{Koehler1986}. The principal cells of layer VI are mainly 
globular and polygonal cells~\cite{Lingenhoehl1991}. 

\subsubsection{Layer II} 

The principal neurons in EC layer~II are of stellate and pyramidal 
morphology. In addition, small numbers of neurons with fusiform, horizontal 
tripolar, and bipolar shapes were found. The distribution of the two main neuron
types, stellate cells and pyramidal cells, is not uniform between LEA and MEA. 
In the latter stellate cells are more abundant and pyramidal cells are mostly 
found near the layer~III border~\cite{Klink1997}. 

\begin{figure}[t]
  \centering
  \includegraphics[width=1.0\textwidth]{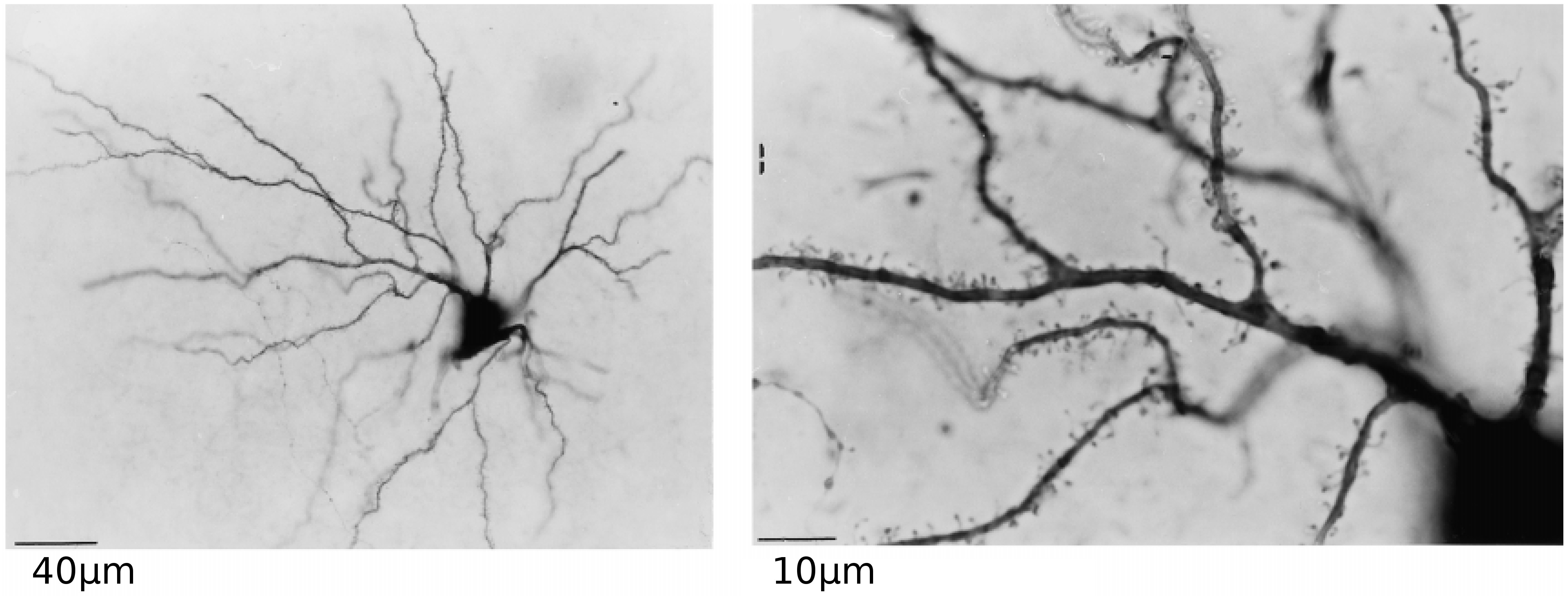}
  \caption[Stellate Cell]{Photomicrograph of a stellate cell in layer~II of the
    MEA. ({\bf{left}}) The numerous primary dendrites taper gently and remain
    relatively thick far from the soma. ({\bf{right}}) Image section showing 
    thick dendrites covered evenly with spines.
    Figure adapted from Klink and Alonso~\cite{Klink1997}.}
  \label{fig:stellate}
\end{figure}

The stellate neurons in layer~II have triangular, rectangular, or trapezoidal
cell bodies~\cite{Lingenhoehl1991}. They possess several thick, primary 
dendrites which decrease in numbers dorsoventrally from an average of ten down 
to seven (max.~14, min.~5)~\cite{Garden2008}. The average ratio of primary
dendrites to the number of dendritic endpoints is about 1:15 (min.~1:10.8, 
max.~1:24.7) and measurements of the overall dendritic length range from 12.8mm
to 18.1mm~\cite{Lingenhoehl1991}. Reports on the shape of the dendritic domain 
vary. Lingenhöhl and Finch~\cite{Lingenhoehl1991} report a spectrum of shapes 
from circular domains with centrally located soma to broad, ellipsoidal domains 
with marginally located soma. Klink and Alonso~\cite{Klink1997} describe double 
V-shaped, bi-triangular dendritic domains. Others refrain from characterizing 
the overall, general shape of the dendritic 
domain~\cite{Germroth1989,Quilichini2010,Burgalossi2014}. In general the apical
dendrites of stellate cells branch profusely in layer~II and layer I and reach
up to the pial 
surface~\cite{Germroth1989,Lingenhoehl1991,Klink1997,Quilichini2010}. The basal 
dendrites extend within layer~II and superficial regions of layer 
III~\cite{Klink1997,Quilichini2010}. In some cases basal dendrites extend down 
to layer IV~\cite{Lingenhoehl1991}. The entire dendritic tree of the stellate 
cells in layer~II is evenly covered by dendritic spines 
(fig.~\ref{fig:stellate}) with an estimated 0.5 to 1~spines per 
1µm~\cite{Lingenhoehl1991}. Considering an overall dendritic length of about 
15mm leads to an estimate of 7500 to 15000 spines per stellate cell where each 
spine can host one or more synaptic connections. The stellate cell's axon gives
rise to about three to five axon collaterals within its first 400µm. The axon 
collaterals are oriented towards the superficial layers, branch repeatedly, and
form a ``delicate net over the entire dendritic domain''~\cite{Klink1997}. A
number of long axons branches extend parallel within layer I in mediolateral
direction beyond the cell's dendritic field~\cite{Klink1997,Quilichini2010}. 
Additionally, axon collaterals spawning in deep portions of layer~III and in 
layers IV to VI were also observed~\cite{Klink1997}.

Pyramidal neurons in layer~II have one or two apical dendrites that start to 
branch at the border between layer~II and layer I. Secondary dendrites appear
to branch off almost perpendicular resulting in a bitufted appearance of the
dendritic arbor~\cite{Germroth1989,Klink1997}. The basal dendrites of layer~II 
pyramidal cells branch extensively around the soma and are confined to layer~II 
and the upper parts of layer~III. The density of dendritic spines is higher in 
comparison to stellate cells, especially in case of the apical 
dendrites~\cite{Klink1997}. The ratio of primary dendrites to the number
of dendritic end points was measured in one case to be 
1:9.4~\cite{Lingenhoehl1991}. The axon of layer~II pyramidal cells is thin and
meanders through layer~II,III, and IV giving off several, mainly horizontal 
collaterals that branch multiple times. Some collaterals ascend to upper 
layers~\cite{Germroth1989,Klink1997}. The main axon continues on a radial path
towards the angular bundle\footnote{The angular bundle lies below layer 
VI.} (AB)~\cite{Klink1997}. 

\subsubsection{Layer III} 

The principal neurons of EC layer~III are pyramidal cells with a prominent, 
triangular cell body and radially extending primary dendrites. The apical 
dendrites of large pyramidal cells bifurcate in layer~III and extend further 
into layer~II and layer I where they continue to branch. Smaller pyramidal cells
do not bifurcate but still extend into layer~II and layer I. Basal dendrites 
branch extensively in layer~III, but extend to deep layers as 
well. All dendrites of layer~III pyramidal cells are densely covered with 
spines~\cite{Lingenhoehl1991,Quilichini2010}. Lingenhöhl and Finch report of up
to five primary dendrites with an average ratio between primary dendrites and 
the number of dendritic endpoints of about 1:12 (min.~1:11.2, 
max.~1:15.5)~\cite{Lingenhoehl1991}. In the same study the total dendritic 
length of a big pyramidal neuron was measured to be 11.3mm and the corresponding 
length of a small pyramidal neuron was measured to be 6.6mm. The axon of layer 
III pyramidal cells gives off several collaterals already within layer~III which
either extend parallel to the layer along the anteposterior axis or extend 
towards the superficial layers where they stay restricted to the region occupied
by the dendritic domain of the cell. Further collaterals of the axon branch off
in layer V. Compared to the axons of layer~II neurons, the axonal branches of
layer~III pyramidal cells appear to be distributed more 
evenly~\cite{Quilichini2010}.

\subsubsection{Layer IV} 

Layer IV of the EC is commonly regarded as virtually devoid of neurons. However,
Lingenhöhl and Finch~\cite{Lingenhoehl1991} managed to sample a number of 
sparsely scattered regular-sized and large-sized pyramidal cells as well as one 
spindle cell in this layer. The larger pyramidal cells and the spindle cell were
located at the border to layer V. The regular-sized pyramidal cells (n = 3) had 
between six and eleven primary dendrites with an average ratio between primary 
dendrites and the number of dendritic endpoints of about 1:9 (min.~1:7.4, 
max.~1:10.1). The total dendritic length of one cell was estimated to be 8.3mm. 
The basal dendrites branched predominantly in layer IV and layer V with a few 
dendrites reaching up to deep portions of layer~III. The apical dendrites 
extended into layer~II and layer I where they ramified. On their way up they 
bifurcated already in layer IV without any further bifurcation in layer~III, 
leaving the latter essentially free of layer IV dendrites. Spines were present 
on all parts of the dendritic tree. The larger pyramidal cells (n = 2) at the 
border to layer V had five and eight primary dendrites, respectively with  
ratios between primary dendrites and the number of dendritic endpoints of 1:11.2
and 1:8.8. The total dendritic length was measured as 13.0mm and 13.3mm. In 
general the dendritic domains were similar to that of the regular-sized 
pyramidal cells. Yet, spines were present only after the first or second 
dendritic bifurcation. The spindle cell had only 3 primary dendrites and a ratio
between primary dendrites and the number of dendritic endpoints of 1:13.6. The
total length of the dendritic domain was estimated to be 8.7mm. A single apical
dendrite extended into layer~II and layer I where it branched extensively. The
two basal dendrites extended towards the subiculum.

\subsubsection{Layer V/VI} 

The principal neurons of EC layer V and layer VI are pyramidal cells with an
average of 5.4 primary dendrites (min.~3, max.~7) and an average ratio between
primary dendrites and the number of dendritic endpoints of 1:5.6 (min.~1:4, 
max.~1:8.7). This ratio is significantly lower compared to the ratios of neurons
in the upper layers~\cite{Lingenhoehl1991}. The basal dendrites distribute 
horizontally in layer V and layer VI where they branch sparsely or not at all. 
In one cell the basal dendrites extended into layer IV. The apical dendrites 
either bifurcate in layer IV and deep layer~III or they bifurcate in layer~III. 
In the former case the dendrites branch extensively in layer~IIa and layer I. 
In the latter case the dendrites branch only sparsely until they reach layer I.
Spines were present on all segments of the dendritic domain. The average 
dendritic length was estimated to be 6.2mm (min.~4.8mm, 
max.~7.2mm)~\cite{Lingenhoehl1991,Quilichini2010}. Axonal branches were evenly
distributed in layer V (66.1\%) as well as in layer~III 
(33.3\%)~\cite{Quilichini2010}.

\subsubsection{Interneurons}

In a recent publication Buetfering et al.\@~\cite{Buetfering2014} were able to 
shed some light on the properties of interneurons in layer~II of the MEA with 
the help of optogenetics. They used an adeno-associated virus to deliver and 
selectively incorporate channelrhodopsin-2, i.e., a light-activated ion channel 
into approximately 50\% of the population of GABAergic\footnote{inhibitory} 
interneurons in the MEA of mice. This enables the external activation of the 
interneurons by locally applying pulses of blue laser light through an optic 
fiber implanted parallel to the common set of tetrodes used for recording the 
electrical activity of the neurons.

The targeted interneurons control the activity of principal neurons in layer~II
by local GABAergic connections. The axons of the interneurons extend and branch 
widely within layer~II where they form basket-like complexes around other 
neurons. Their dendritic trees are mainly located in layer I and only sparsely 
covered with spines receiving excitatory (fast AMPA-mediated and slow 
NMDA-mediated) as well as inhibitory (GABAergic) input \cite{Jones1993}. A part
of this input comes from grid cells with various phases. As a consequence the
firing pattern of interneurons is not grid-like. Like many other cells of the
MEA the firing rate of interneurons is modulated by running 
speed~\cite{Buetfering2014}. 

Collective activation of the interneurons via the light-activated ion channels
silenced neurons of all cell types reliably. The delay (\(<\) 5ms) between the onset
of the laser pulse and the following inhibition of postsynaptic cells is 
indicative of a monosynaptic connection between interneurons and principal 
cells. The recovery of inhibited, postsynaptic cells after the laser was 
switched off took 25ms. This duration is consistent with the dynamic properties
of \(\mathrm{GABA}_A\) receptors~\cite{Buetfering2014}.

Based on their results Buetfering et al.\@~\cite{Buetfering2014} speculate that 
interneurons may control the gain of the grid cells rather than being essential 
for the generation of the grid pattern.

\subsection{Topographical Organization and Microcircuits} 

The topographical and modular organization (section~\ref{sec:toporg}) of grid cells
on a functional level raises the question whether this organization is reflected
in the underlying neuronal structure. Within the last three decades a number of 
studies~\cite{Koehler1986,Ikeda1989,Lingenhoehl1991,Hevner1992,Witter2006,Garden2008,Quilichini2010,Burgalossi2011,Bonnevie2013,Burgalossi2014,Gu2018}
have identified several anatomical properties of neurons in the MEA that exhibit
a dorsoventral gradient and/or a modular organization. However, it is yet 
unknown if these anatomical properties are causally related to the observed
functional organization of grid cells~\cite{Burgalossi2014}. A very recent study
by Gu et al.\@~\cite{Gu2018} provides first tentative evidence that grid cells 
with similar grid phases form local clusters in the MEC that are arranged in a
noisy two-dimensional lattice. 

Many of the anatomical properties reported below were identified by 
histochemical methods, i.e., the selective staining of cell parts by exploiting
unique chemical properties of those parts~\cite{Lang2013}. This includes direct 
binding of dyes to negatively charged nucleic acids (e.g., Nissl staining using 
cresyl violet~\cite{Burgalossi2011}), incorporating dyes in the process of 
enzyme catalysis (e.g., cytochrome 
oxidase~\cite{WongRiley1989,Hevner1992,Burgalossi2011}), or utilizing the 
selectivity of exogenous antibodies towards antigens present in the cells (e.g.,
reelin immunoreactivity~\cite{Burgalossi2014}).

\begin{figure}[t]
    \begin{center}
        \subfigure[]{\includegraphics[width=0.34\textwidth]{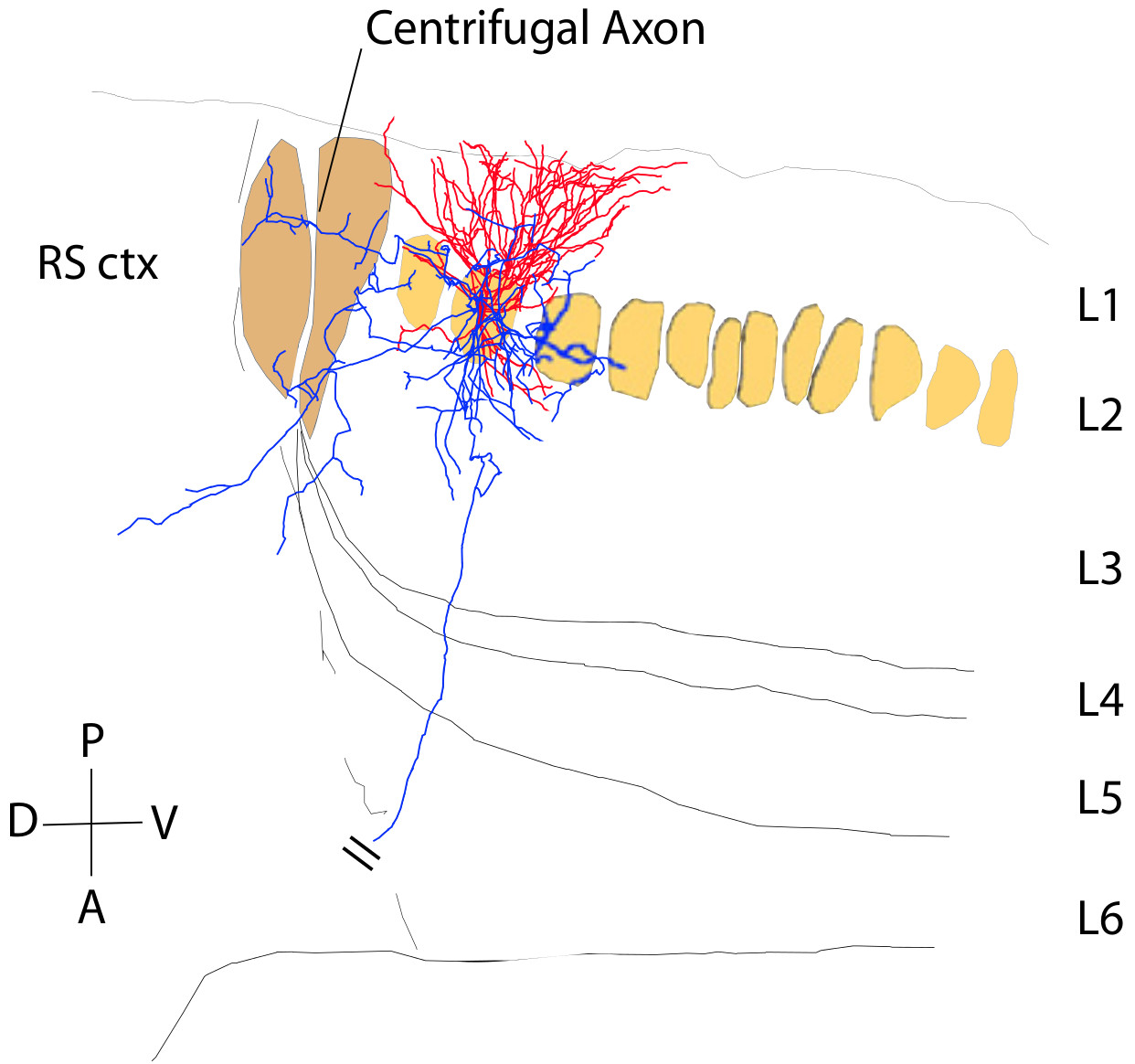}}\hspace{1.0cm}
        \subfigure[]{\includegraphics[width=0.56\textwidth]{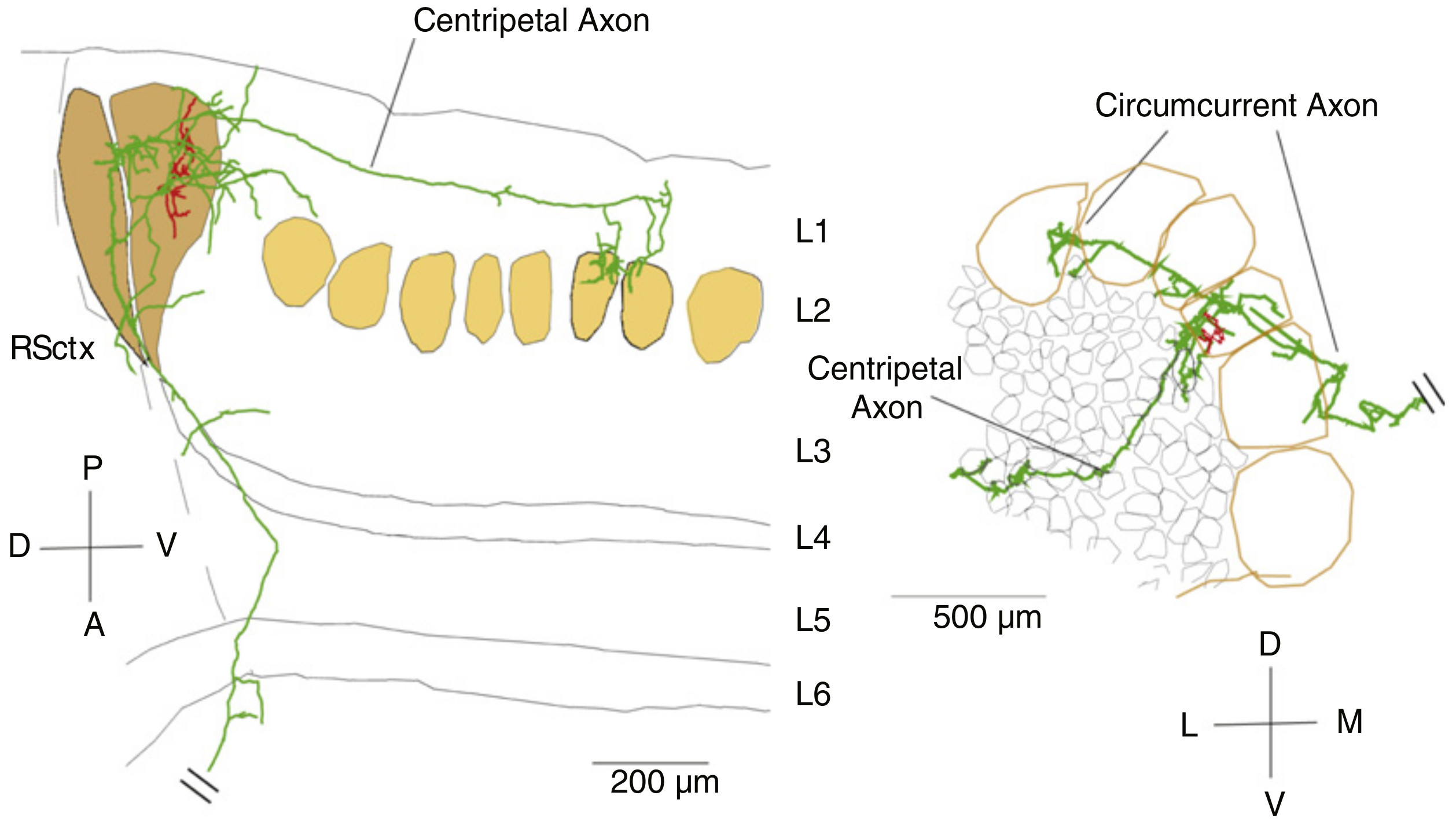}}
        \caption[Stellate Cell Microcircuit]{Two reconstructions of neuronal 
          morphologies in the MEA. ({\bf{a}})~Reconstruction of axons (blue) and
          dendrites (red) of a layer~II stellate cell. Light brown regions 
          indicate small layer~II patches of cytochrome oxidase activity. Dark 
          brown regions indicate large patches of cytochrome oxidase activity at
          the border of the MEA. D:~dorsal, V:~ventral, P:~posterior, 
          A:~anterior, RS~ctx:~retrosplenial cortex. ({\bf{b}})~Reconstruction 
          of axons (green) and dendrites (red) of a cell located in a border 
          patch identified by cytochrome oxidase activity. Left: parasagittal
          section, right: tangential section. L:~lateral, M:~medial, other 
          labels as in (a).
          Figure adapted from Burgalossi et al.\@~\cite{Burgalossi2011}.}
        \label{fig:stelDet}
    \end{center}
\end{figure}

\begin{figure}[t]
    \begin{center}
        \subfigure[]{\includegraphics[width=0.6\textwidth]{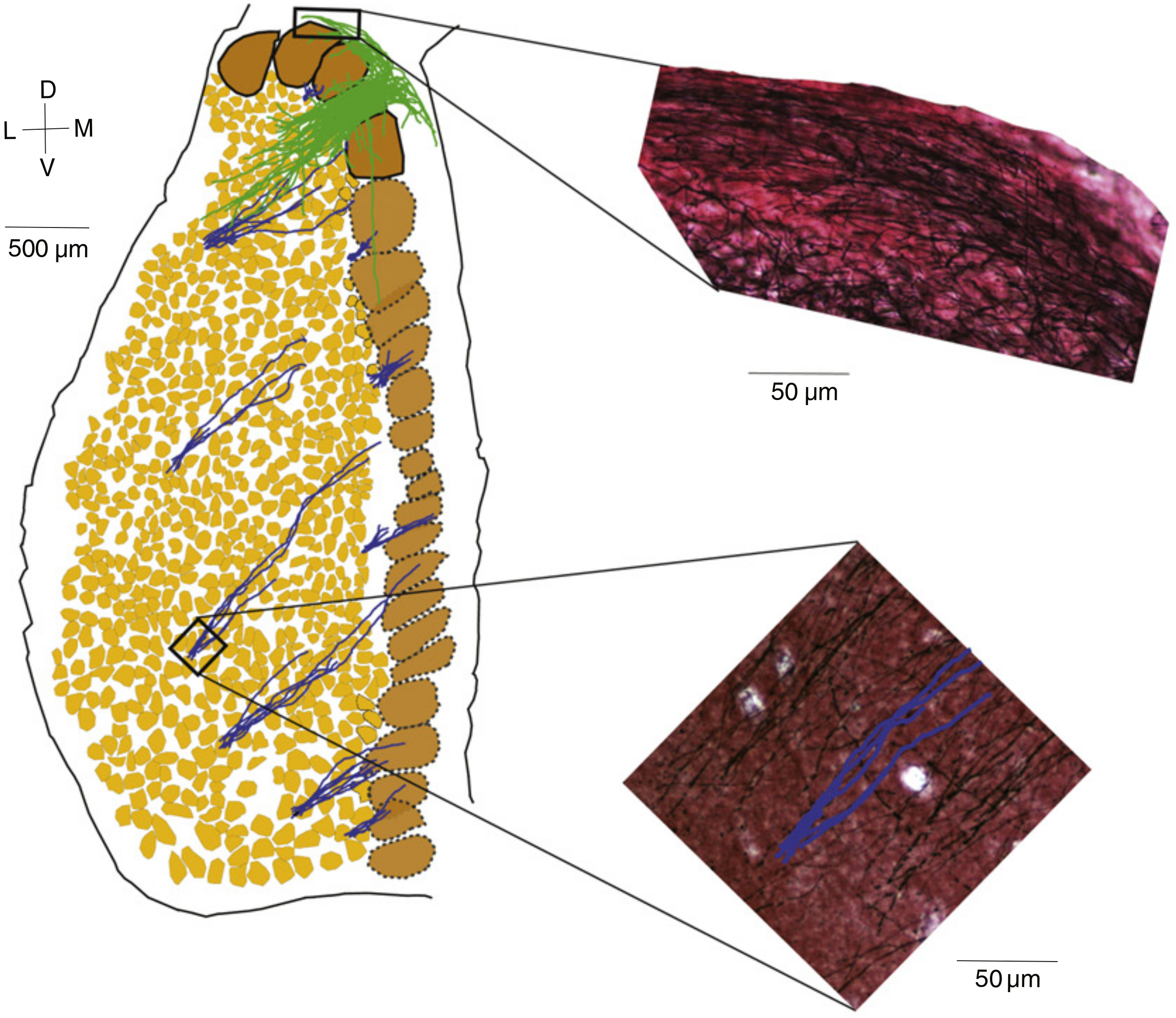}}\hspace{1.0cm}
        \subfigure[]{\raisebox{2cm}{\includegraphics[width=0.3\textwidth]{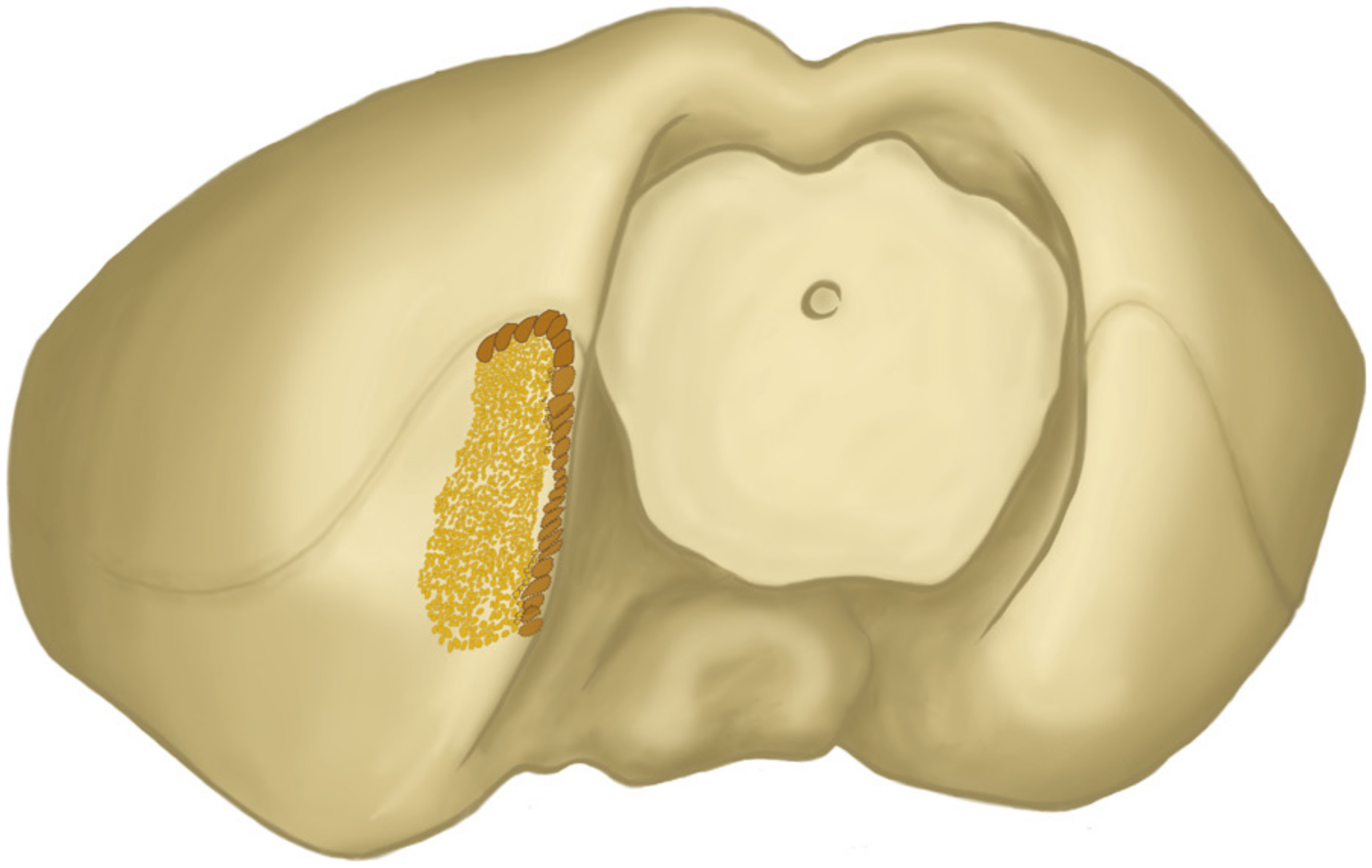}}}
        \caption[MEA Layer I Axons]{Large-scale overview of MEA layer I axons. 
          ({\bf{a}})~Reconstruction of MEA based on tangential sections. The 
          sections were stained for myelin. Light brown patches are cell 
          clusters in layer~II identified in the sections as clusters of white 
          somata surrounded by myelin. Dark brown patches are identified 
          indirectly by extensive axon bundles on the dorsal surface of each
          patch. Bundles of putative centrifugal axons (blue) densely populate
          layer I with about twice as many bundles as small patches. Axons of 
          cells within one large patch are drawn in green. D:~dorsal, 
          V:~ventral, L:~lateral, M:~medial. ({\bf{b}}). Reconstruction of the
          MEA patches superimposed on a posteriorlateral view of the rat brain.
          Figure adapted from Burgalossi et al.\@~\cite{Burgalossi2011}.}
        \label{fig:patchDet}
    \end{center}
\end{figure}

The layers of the MEA contain regularly distributed clusters of neurons, which
are sometimes referred to as {\em islands}. Staining for the enzymes glycogen 
phosphorylase and cytochrome oxidase reveals these clusters in layer~I and 
layer~III as well as in layer~II (fig.~\ref{fig:stelDet}), respectively. Both
enzymes are part of metabolic processes and the intensity of the resulting stain
reflects the metabolic activity in the particular 
region~\cite{Hevner1992,Burgalossi2011,Burgalossi2014}. The presence of cell
clusters in layer~II is also indicated by staining for myelin 
(fig.~\ref{fig:patchDet})~\cite{Burgalossi2011} and by staining for 
immunoreactivity to the calcium-binding phosphoprotein R2D5. In the latter case 
layer~III cells are predominantly R2D5 positive while layer~II cells are R2D5 
negative resulting in a pattern of interweaved, separate 
columns~\cite{Ikeda1989}. Within these columns the dendritic fields of layer~II
and layer~III neurons receive separate projections from different brain 
areas~\cite{Hevner1992}. The cell clusters in layer~II are locally uniform in 
size and are surrounded by myelinated fibers. Along the dorsoventral axis the 
average size of the clusters increases while the average cell size and the 
degree of myelination decreases. Adjacent to these cell clusters lies a series 
of larger clusters along the dorsal and medial borders of the 
MEA (fig.~\ref{fig:stelDet} and~\ref{fig:patchDet})~\cite{Burgalossi2011}.

\begin{figure}[t]
  \centering
  \includegraphics[width=1.0\textwidth]{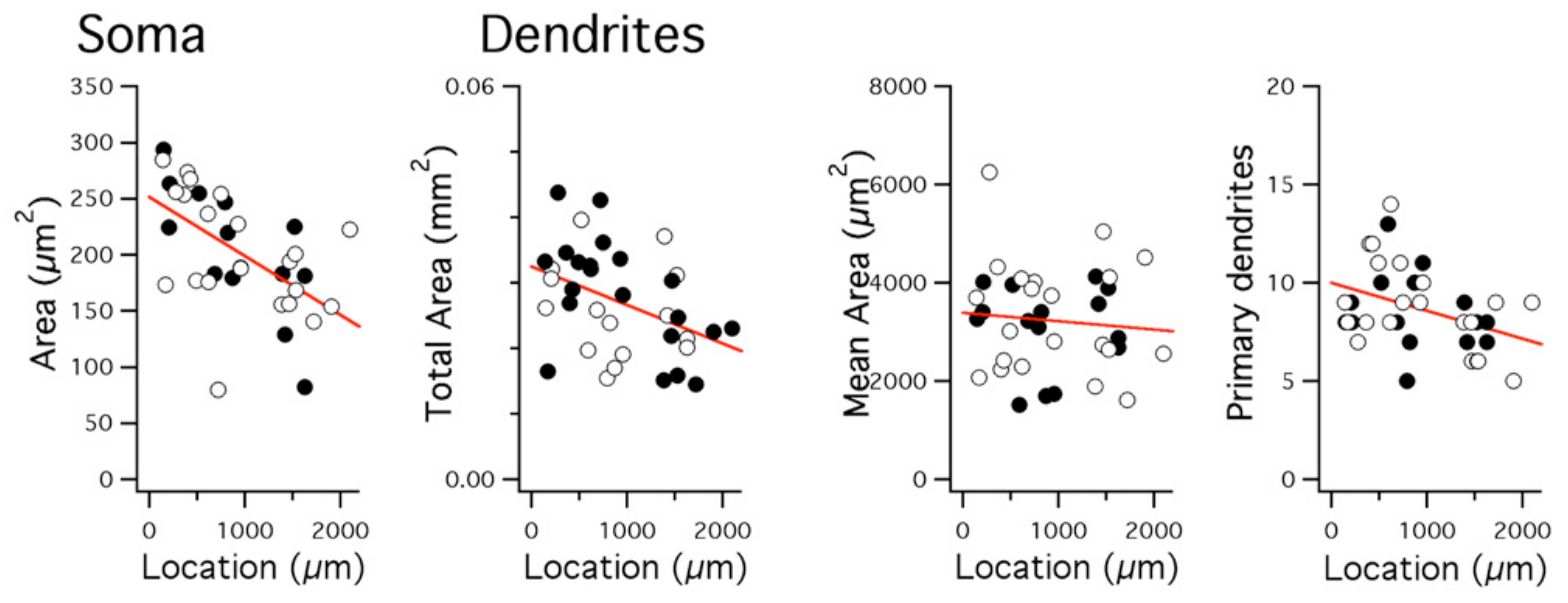}
  \caption[Stellate Cell Properties]{Variation of morphological properties of 
    MEA layer II stellate cells along the dorsoventral axis. The plots show
    (from left to right) the cross sectional soma area, total dendritic surface
    area, average dendritic surface area, and number of primary dendrites of
    reconstructed stellate cells plotted against the soma's distance from the 
    dorsal pole. Open and closed circles refer to two different 
    recording conditions used to measure the cell's electrotonic properties.
    Both conditions did not influence the measurement of the morphological 
    properties shown here.
    Figure adapted from Garden et al.\@~\cite{Garden2008}.}
  \label{fig:stellate_nod}
\end{figure}

The variation of cluster size, cell size, and degree of myelination along the
dorsoventral axis is accompanied by further variations on the level of 
individual cell properties. Garden et al.\@~\cite{Garden2008} studied these 
variations in MEA layer II stellate cells in detail. Morphologically layer II
stellate cells show a dorsoventral decrease in cell body perimeter, cell body
cross sectional area, and dendritic surface area. The latter is not caused by
any changes to the surface area of individual dendritic branches, but by an
overall reduction in the number of primary dendrites from an average of ten 
primary dendrites at the dorsal pole to an average of seven primary dendrites
at more ventral locations (fig.~\ref{fig:stellate_nod}). Furthermore, the 
total number of dendritic branch points also decreases dorsoventrally. In 
addition to these morphological changes the electrotonic properties of layer II
stellate cells change too along the dorsoventral axis. In particular, the input
resistance and the membrane time constant increase dorsoventrally. As a 
consequence the current threshold to trigger action potentials (APs) decreases 
4-fold along the dorsoventral axis, i.e., the amplitude of positive current 
required to trigger an AP is much larger at dorsal neuron locations. Moreover,
the time window for the detection of coincident inputs also varies in 
dorsoventral direction. It is about three times wider in the most ventral 
locations than it is in the most dorsal ones.

Information about the microcircuitry within the EC is incomplete and commonly 
based just on information about the coarse morphology of principal neurons. 
Potential connections between neurons are often extrapolated based on 
overlapping input and output regions which are deduced from the neurons' 
dendritic and axonal domains. Although such piecewise and tentative information 
does only provide limited insight into the actual neuronal circuitry it does 
provide a set of constraints to which possible models and explanations of grid 
cell behavior have to adhere. To this end, the following paragraphs summarize 
the available, but rather fragmentary information about EC microcircuitry.

The intra-EC projections of principal neurons can be described by their 
horizontal extent parallel to the layers of the EC and their longitudinal extent
in the orthogonal direction. In general, neurons in layer~III are more 
restricted in their horizontal and longitudinal extent than neurons in layer~II.
Together they are both more restricted in their extent than neurons in layers~IV
to VI~\cite{Koehler1986}. In all layers, the majority of projections are 
oriented longitudinally and distribute between the respective layer of origin
and the layers above, i.e., cells in layer~V project to layers~V, IV, and III;
cells in layer~III project to layers~III, II, and I; and cells in layer~II 
project within layer~II as well as adjacent parts of layers~I and 
III~\cite{Witter2006}. In addition, minor projections in the opposite direction
were also observed, i.e., from layer~II cells to layers~IV, V, and 
VI~\cite{Klink1997,Witter2006}; and from layer~III cells to 
layer~V~\cite{Quilichini2010}. By comparison layer~V neurons contribute three 
times more intrinsic connections than neurons in layer~II and five times more 
than neurons in layer~III. The axonal arbor of layer~V neurons is cone shaped 
with its base in layer~V and its peak in layer~II. This suggests that layer~V 
cells predominantly interact with each other via their basal dendrites. In 
contrast, the axonal domain of layer~II neurons forms an inverted cone with its 
base in layer~I and its peak between layers~III and V. This allows layer~II 
cells to communicate with a wide range of neurons in layers~II, III, and V. 
Layer III neurons may represent some form of bidirectional link between deep and
superficial layers as they can receive input from all layers and possess axons 
that converge on smaller subsets of layer~II and/or layer~V 
cells~\cite{Quilichini2010}. Local excitatory connections between principal
cells were observed in layer~III~\cite{Winterer2017,Witter2006} and 
layer~V~\cite{Witter2006}, but none~\cite{Witter2006,Burgalossi2014} or 
few~\cite{Quilichini2010} in layer~II. The latter results are called into 
question in a recent study by Winterer et al.\@~\cite{Winterer2017} who measured
(in vitro) intra-layer excitatory connections between layer~II stellate cells at
a rate of \(2.5\%\). In the same study Winterer et al.\@ also report on 
predominantly unidirectional excitatory connections between layer~II/III 
pyramidal cells and layer~II stellate cells (\(13.5\%\) and \(7.0\%\), 
respectively).

Neurons in the larger clusters at the EC border are targeted by axons of layer 
II stellate cells (fig.~\ref{fig:stelDet}a) as well as layer~III pyramidal 
cells. The so called {\em centrifugal axons} appear to be bundled and oriented 
in dorsomedial direction (fig.~\ref{fig:patchDet}a). The targeted cells in the 
large clusters differ in their morphology from all other neurons in the MEA. 
Their dendritic trees are small and extend not beyond their home cluster. They 
possess three main axons: one axon descending towards the presubiculum, one {\em 
circumcurrent axon} targeting many other large clusters along the EC border, and
one {\em centripetal axon} specifically targeting one to two small layer~II 
clusters in which it arborizes. In addition, a number of axons branch locally 
within the home cluster (fig.~\ref{fig:stelDet}b). The firing characteristics of
large cluster cells range from spatially multi-peaked firing to spatially broad 
tuning, all with a high degree of modulation by head 
direction~\cite{Burgalossi2011}. 

A more detailed view on the microcircuitry of MEA layer~II was recently provided
by Varga et al.\@~\cite{Varga2010}. They identified two major, non-overlapping
cell groups in layer~II by utilizing the cells' immunoreactivity to either 
reelin (\(53\pm2.6\%\) of all cells) or calbindin (\(44\pm2.2\%\) of all cells).
Only \(2.8\pm1.1\%\) of all cells were reelin and calbindin double 
positive\footnote{Total number of cells analyzed in 3 rats: 1152}. More 
importantly, injection of the retrograde tracer biotinylated dextrane amine 
(BDA) into the ipsilateral dentate gyrus labeled predominantly (\(98.5\pm0.5\%\)) 
the reelin positive principal cells indicating that only these cells project to
the dentate gyrus. In contrast, the calbindin-expressing principal cells were
found to project extra-hippocampally to the contralateral entorhinal cortex. In
addition, Varga et al.\@ were able to show that reelin positive and calbindin 
positive cells each interact with a different population of inhibitory 
interneurons. Reelin-expressing principal cells interact with fast-spiking 
interneurons, whereas calbindin-expressing cells interact with interneurons 
that form basket-like axonal domains surrounding the principle cells providing
perisomatic inhibition. The observations of Varga et al.\@ are confirmed by a
more recent study by Ray et al.\@~\cite{Ray2014}. In addition, Ray et al.\@ were
able to show that calbindin negative cells were primarily stellate cells whereas
calbindin positive cells were primarily pyramidal cells. The latter were 
observed to form hexagonal patches near the layer I/II border. This hexagonal
structure led Ray et al.\@ to hypothesize that the cells' arrangement ``might be
an isomorphism to hexagonal grid activity''. However, they were not able to 
measure the spatial modulation in a sufficient number of identified cells to 
assess the validity of this hypothesis. In a related study Tang et 
al.\@~\cite{Tang2014} were able to find potential evidence that grid cells are 
``preferentially recruited'' from the population of calbindin positive cells, 
i.e., pyramidal cells, and that border cells stem preferentially from  
calbindin negative cells, i.e., stellate cells.

Regarding projections into the EC Lingenhöhl and Finch~\cite{Lingenhoehl1991}
as well as Bonnevie et al.\@~\cite{Bonnevie2013} could make interesting, and to
some degree contradicting observations. Lingenhöhl and 
Finch~\cite{Lingenhoehl1991} investigated projections from the hippocampal area 
to the EC in vitro. In response to electrical stimulation of either the DG, CA1 
or CA3 all cells that were observed in the EC showed synaptic responses in form
of inhibitory postsynaptic potentials (IPSPs). No clear excitatory postsynaptic
potentials (EPSPs) could be identified indicating that projections from 
hippocampal areas to the EC are predominantly inhibitory. However, Bonnevie et 
al.~\cite{Bonnevie2013} investigated projections from the hippocampus to the MEA
in vivo on a functional level. They observed grid cells in the MEA before and
after the inactivation of the hippocampus. In the latter case grid cells lost
their grid-like firing pattern and became direction sensitive. Based on these
results Bonnevie et al.\@ concluded that grid cells require excitatory drive
from the hippocampus. The results of Lingenhöhl and Finch and Bonnevie et al.\@
do not necessarily contradict each other. The postulated excitatory drive could 
reach the EC on an indirect route, e.g., via the subiculum or the pre- and 
parasubiculum.

\section{Conclusions}
\label{sec:summary}

Since their discovery a decade ago grid cells have attracted considerable 
attention resulting in a better and more detailed understanding of their 
functional characteristics as well as a better understanding of the 
para\-hip\-po\-campal-hippocampal region their embedded in. The observation of
new grid cell phenomena like the strong influence of environmental geometry or
the fragmentation of grid patterns question early hypothesis on grid cell 
function and call for revised models of grid cell behavior that allow to 
formulate new hypothesis and direct further experimental investigation. 
Considering the presence of related cell types like border cells, head 
direction cells, irregular spatial cells, and non-spatial cells that all 
converge on hippocampal neurons and may receive feedback from these in turn, it 
appears essential that new models should aim to integrate these different
types of signals into a single, conceptual framework that explains the 
processing of information within a wider context that extends beyond 
isolated descriptions of individual cell types and their behavior.

\bibliographystyle{plain}
\bibliography{kerdels_peters_2018_grid_cell_survey}

\end{document}